\documentclass[aps,prd,showpacs,preprintnumbers,amsmath,amssymb,superscriptaddress,floatfix,nofootinbib,10pt]{revtex4-1}

\usepackage{graphicx}
\usepackage{epsfig}
\usepackage{epstopdf}
\usepackage{hyperref}
\usepackage{amsmath}
\allowdisplaybreaks[1]
\usepackage{amsfonts}
\usepackage{amssymb}
\usepackage{multirow}
\usepackage{makecell}
\usepackage[dvipsnames]{xcolor}
\usepackage{ulem}
\usepackage{array}

\usepackage{mathtools}
\usepackage{bm}
\usepackage{caption}
\usepackage{subcaption}
\usepackage{booktabs}
\usepackage{float}
\usepackage{slashed}
\usepackage{gensymb}
\usepackage{makecell} 

\newcommand{\UCAS}{School of Physical Sciences, University of Chinese Academy of \\ Sciences (UCAS), Beijing 100049, China}
\newcommand{\ZZU}{School of Physics and Microelectronics, Zhengzhou University, Zhengzhou, Henan 450001, China }

\newcommand{\ud}{\mathop{}\negthinspace\mathrm{d}}  
\newcommand{\normofvector}[1]{\left|\boldsymbol{#1}\right|}  
\newcommand{\abs}[1]{\left|#1\right|} 
 
\newcommand{\gammaL}{\gamma_{\mathrm{L}}}
\newcommand{\betaL}{\beta_{\mathrm{L}}}

\setcellgapes{2pt} 

\makeatletter
\g@addto@macro\normalsize{
  \setlength\abovedisplayskip{20pt}
  \setlength\belowdisplayskip{20pt}
  \setlength\abovedisplayshortskip{20pt}
  \setlength\belowdisplayshortskip{20pt}
}
\makeatother

\begin{document}
\title{Research on the Phase space of three- and four-body final states process}

\author{Kang Yu}
\affiliation{\ZZU}
\affiliation{\UCAS}
\author{De-Min Li}\email{lidm@zzu.edu.cn}
\affiliation{\ZZU}
\author{Jia-Jun Wu}\email{wujiajun@ucas.ac.cn}
\affiliation{\UCAS}

\begin{abstract}
The analytical formulae for the phase space factors and the three-momenta of three- and four-body final states are derived for all sets of independent kinematic variables containing invariant mass variables.
These formulae will help experimental physicist to perform the data analysis.
As an example, we show how to use these formulae to distinguish the different mechanisms of $e+p\to e+J/\psi+ p$ process for searching the signals of $P_c$ states at the energy region of  Electron-Ion collider at China (EicC). 
\end{abstract}

\maketitle

%
%
\section{Introduction}
\label{sec1}

%
Internal structure and interaction mechanism of microscopic particles is one of the main issues in the field of particle physics. 
However, due to the extremely short reaction time, the intermediate states of reaction process can not be measured directly until now.
Nevertheless, the distribution of final states, which can reflect the internal structure as well as interaction mechanism, are measurable. 
Through these distributions, physicists are able to explore the nature of the various particles and their internal structures.
For instance, in 1911 Ernest Rutherford revealed the internal structure of atomic by analyzing the angular distribution of outgoing particles in the well-known gold foil experiment.  
Therefore, the differential cross-section and the differential decay width play an important role in studying particle physics.
%

%
From the Review of Particle Physics (RPP) ~\cite{ParticleDataGroup:2020ssz}, the differential cross-section for the $2\to n$ scattering process and differential decay width of a particle into $n$ bodies can be written as follows, respectively, 
\begin{align}
        \ud\sigma &= \frac{(2\pi)^4}{4\sqrt{(q_{1}\cdot q_{2})^2-m_1^2m_2^2}}|\mathcal{M}|^2 \ud\Phi_n\,,
        \label{eq:dsigma}
        \\
        \ud\Gamma &= \frac{(2\pi)^4}{2m}|\mathcal{M}|^2 \ud\Phi_n\,,
        \label{eq:dGamma}
    \end{align}
where $q_i$ and $m_i(i=1,2)$ are the four-momentum and mass of $i$-th initial particle in the scattering process, respectively, and $m$ is the mass of parent particle in the decay process.
$\mathcal{M}$, which depends on the dynamic mechanisms, is the Lorentz invariant amplitude and
$\ud\Phi_n$  named as phase space is a purely kinematic factor which is conventionally defined in the following Lorentz-invariant form 
\begin{equation}
    \ud\Phi_n = \delta^{(4)}\left(P - \sum\limits_{i=1}^n p_i\right)\prod\limits_{i=1}^n \frac{\ud^3\boldsymbol{p}_i}{\left(2\pi\right)^3 2 E_i}\,, 
    \label{eq:dphi}
\end{equation} 
where $P$ is the summation of the four-momenta of all initial states, and $p_i=(E_i, \boldsymbol{p}_i)$ is the four-momentum of the $i$-th particle in the final states.
The phase space actually makes a bridge between the theoretical calculation for $\mathcal{M}$ and the experimental observation for $\ud\sigma$ or $\ud\Gamma$.
The computation of the phase space is of great significance for experimental physicists to analyze distribution data and extract theoretical variables.
%


One of the most important task for particle physicists is to extract the resonance from the invariant mass spectrum of the final states.
Generally, besides the invariant amplitude $\mathcal{M}$, the phase space factor also plays an important role in $\ud\sigma$ and $\ud\Gamma$ in the invariant mass spectrum. 
Therefore, it is necessary to express the phase space factor in terms of the various invariant masses variables.
For example, in the chapter of Kinematics in the RPP, the three-body phase space is expressed in two forms. 
One contains one invariant mass variable and the other two independent invariant masses which can be visualized by the well-known Dalitz plot.
There have been some work on the phase space of $n$-body final states with $n>3$. 
Some systematic methods are introduced in the textbooks and literatures, for example, Refs.\cite{byckling1973particle, PhysRev.185.1865, PhysRevD.2.1902, POON1970509, Gehrmann-DeRidder:2003pne}.
All of these works provide the various formulas to calculate three-, four- and n-body phase space distribution and integration.
Recently, in Ref.~\cite{Jing:2020tth}, a new systematic graphic method to decompose an arbitrary $n$-body phase space is introduced.

In the experimental side, more and more new particles are discovered from the three- and four-body final states.
If the invariant masses variables are properly chosen, the resonance can be extracted much more efficiently. 
Otherwise, the signal is not obvious and sometimes may even be buried in the background. 
Therefore, in this paper, we focus on the three- and four-body final states and present the expressions of $\ud\Phi_3$ and $\ud\Phi_4$ directly in terms of all possible sets of invariant mass variables by using a new formulation, which can be served as a handbook which may be convenient as well as helpful for the experimental data analysis.

For $n$-body final states, there are $3n$ kinematic variables but only $3n-4$ of them are independent because of the law of energy-momentum conservation.
Therefore, there are 8 independent kinematic variables (IKVs)  for four-body final states.
Particularly, if the system is rotation-invariant, such as a decay process of a non-polarized parent particle, three kinematic variables describing the absolute direction of three-momenta of final particles can be trivially integrated out.
Even so, there are still 5 IKVs for four-body final states.
In this paper, all cases for choosing IKVs within invariant mass variables are listed.
Then the phase space factor is calculated for each case and furthermore, the four-momenta of the four final states are expressed as functions of IKVs.
Once it is done, the amplitude $\mathcal{M}$ of any interaction mechanism can be expressed quite straightforward.
%


This paper is organized as follows. 
After the introduction, the notation of this paper is defined in Section \ref{sec2}.
In the Sections \ref{sec3} and \ref{sec4}, formulae of the phase spaces of three- and four-body final states are enumerated, respectively.
Then by using the formulae given in Section \ref{sec3}, two possible mechanisms of reaction $e + p\to e+J/\psi+p$ at Electron-Ion collider at China(EicC), which will be helpful to search $P_c$ resonance states there, are distinguished.
The related results are shown in Section \ref{sec5}.
Furthermore, we also give an example of four-body case in Section \ref{sec6}.
At last, brief summary is given in Section \ref{sec7}.

\section{Formalism}
\label{sec2}

In this section, some notations used in this paper are introduced.
The main task of this paper is to present all possible phase space factors in terms of different IKVs for three- and four-body final states.
The key problem is how to find all sets of IKVs.
In principle, IKVs can be divided into two parts, angular variables and the others which can be expressed as functions of several invariant mass variables, such as energies of particles.
As discussed before, since the invariant mass spectrum plays an important role in extracting resonance, invariant mass and angular variables are chosen as IKVs in this paper for the further application.

There are two rules which are useful for classifying the different sets of IKVs. 
Firstly, the number of invariant mass variables appearing in the set of IKVs is counted for the preliminary classification.
For example, in the three-body final states, there are only three cases, two, one, and none invariant mass variables in the set of IKVs.  
Secondly, we consider the different patterns of the set of IKVs but do not distinguish the order of particles.
For example, if only two invariant mass variables are in the set of IKVs for the three-body final states, there are three choices as $(m_{12},\,m_{13})$,  $(m_{12},\,m_{23})$, and  $(m_{23},\,m_{13})$, which are all equivalent.
By following the above two rules, there are only three different sets of IKVs in the three-body final states as shown in the next section. 
%
%
However, with regard to the four-body system, it is much more complicated and a new concept of \textit{distribution number} (DN) will be introduced in detail in Section \ref{sec4}.

On the other hand, all the angular variables can be distinguished as three classes: three Euler angles for the whole reaction system, the polar angles in the sub-system, and various angles between the three-momenta of certain two particles.
Firstly, Euler angles $\alpha,\beta,\gamma$ describe the absolute direction in the fixed frame $Op_xp_yp_z$ or equivalently $Oxyz$.
Euler angles here are defined in the $y$-convention.
Assuming that at the beginning, the direction of $\boldsymbol{p_1}$ is along $\boldsymbol{e_z}$ and $\boldsymbol{p_2}$ lies in $p_zOp_x$ plane with $\boldsymbol{p_2}\cdot\boldsymbol{e_x}>0$ and rotating the configuration of momenta around axis of $\boldsymbol{e_z}$, $\boldsymbol{e_y}$ and $\boldsymbol{p_1}$ in succession by $\alpha$, $\beta$ and $\gamma$ respectively, one can obtain the direction of the momenta of final states.
The overall effect of the successive rotations defined above can be described by the matrix as follows, 
\begin{align}
    \mathcal{R} = 
     \begin{pmatrix*}
        \cos \alpha & -\sin \alpha & 0 \\
        \sin \alpha & \cos \alpha & 0 \\
        0 & 0 & 1
    \end{pmatrix*}
    \begin{pmatrix*}
        \cos\beta & 0 & \sin\beta \\
        0 & 1 & 0 \\
        -\sin\beta & 0 & \cos \beta
    \end{pmatrix*}
    \begin{pmatrix*}
        \cos \gamma & -\sin \gamma & 0 \\
        \sin \gamma & \cos \gamma & 0 \\
        0 & 0 & 1 
    \end{pmatrix*}.\label{eq:R}
\end{align}
%
Secondly, when it comes to the rest frame of the composite particle-$i_1i_2...i_m$ with three-momentum $\boldsymbol{p}=\boldsymbol{p_{i_1}}+\boldsymbol{p_{i_2}}+..+\boldsymbol{p_{i_m}}$, its coordinate axes $O p_x^\star p_y^\star p_z^\star$ are built according to the following procedure.
Firstly, $p_z^\star$ axis is chosen to be along the opposite direction of $\boldsymbol{p}$. 
Secondly, $p_y^\star$ axis is defined by $\boldsymbol{e}_y^\star = \boldsymbol{e}_z \times \boldsymbol{e}_z^\star$. 
Thirdly, $p_x^\star$ axis is naturally determined since $Op_x^\star p_y^\star p_z^\star$ is supposed to be right-handed.
Then, the polar angle of the particle in the coordinate $O p_x^\star p_y^\star p_z^\star$ in this paper can be defined unambiguously.
After all IKVs are fixed, the phase space can be expressed as follows,
\begin{align}
    \ud\Phi_n  = A\, \ud m_a \ud m_b \cdots \ud \alpha_1 \ud\alpha_2 \cdots
    \label{eq:FORA}
\end{align}
where $(m_a,\,m_b,\,\cdots)$ and $(\alpha_1, \,\alpha_2 ,\,\cdots)$ indicate invariant mass and angular variables, respectively.
The $A$ which is the phase space factor with the fixed $(m_a,\,m_b,\,\cdots,\,\alpha_1, \,\alpha_2 ,\,\cdots)$ needs to be derived.
Writing down the amplitude $\mathcal{M}$ as a function of IKVs is also meaningful.
Since $\mathcal{M}$ is actually a function of the three-momenta of final states, it can be obtained quite straightforward once the three-momenta can be expressed in terms of IKVs exactly.
Therefore, another task of this paper is to provide explicit formulae with the IKVs.
Such expressions can be quite complicated, so several intermediate variables will be used for the sake of simplification.

In summary, all cases of IKVs with the invariant mass and angular variables for three- and four-body systems will be listed.
Not only the phase space factor $A$ defined in Eq.(\ref{eq:FORA}) but also the explicit expressions of the three-momenta of final states are to be given.

\section{The Phase Space for Three-body final states}
\label{sec3}

There are three distinct sets of IKVs for three-body final states, which contains two, one, and none invariant mass variables, respectively.
In Tables \ref{tab:3a}-\ref{tab:3c}, IKVs, the phase space factor $A$ defined in  Eq.(\ref{eq:FORA}) and the three-momenta of final states are listed for these three cases.
Besides, the three-momentum of the third particle can be obtained by $-\boldsymbol{p}_1-\boldsymbol{p}_2$ and hence will not be shown in the tables.
For the last case shown in Table \ref{tab:3c}, there is no invariant mass variables and $\theta_{ij}$ is the angle between the three-momenta of the $i-$th and the $j-$th particles.
Furthermore, the $|\boldsymbol{p_i}|$ satisfies an equation as shown in the last row of Table \ref{tab:3c}.   
Though the analytical solution exists, the explicit expression is so complicated that will not be shown there.

\begin{table}[t]
    \centering
	\makegapedcells
        \begin{tabular}{|c|c|}
        \hline
        \textbf{IKVs} & 
        $m_{13}^2,m_{23}^2,\alpha,\cos\beta,\gamma$
        \\
        \hline
        \textbf{$A$} & 
		$\frac{1}{8\left(2\pi\right)^9 4 m^2}$
        \\
        \hline
        \textbf{$\boldsymbol{p_{1,\,2}$}} &
        $
        \begin{pmatrix*}
            p_{1x}\\
            p_{1y}\\
            p_{1z}\\
        \end{pmatrix*}
        = \mathcal{R}
        \begin{pmatrix*}
            0\\
            0\\
            \normofvector{p_1}\\
        \end{pmatrix*}
        $ ,
        $\begin{pmatrix*}
            p_{2x}\\
            p_{2y}\\
            p_{2z}\\
        \end{pmatrix*}
        =
        \mathcal{R}
        \begin{pmatrix*}
            \normofvector{p_2}\sin\theta\\
            0\\
            \normofvector{p_2}\cos\theta\\
        \end{pmatrix*}$
        \\
        \hline
        \textbf{}
        &
            $\begin{aligned}
                \normofvector{p_1} &= \frac{\lambda^{\frac12}(m^2,m^2_1,m^2_{23})}{2m}
                \\
                \normofvector{p_2} &= \frac{\lambda^{\frac12}(m^2,m^2_2,m^2_{13})}{2m}
                \\
                \cos\theta &= \frac{2E_1E_2 -(m^2+m_3^2- m_{13}^2-m_{23}^2)}{2\normofvector{p_1}\normofvector{p_2}}   
            \end{aligned}$
        \\
        \hline
        \end{tabular}
    \caption{
    This set of IKVs contains two invariant mass variables.
    ${A}$ is the phase space factor and here 
$\ud\Phi_3 = A \ud m_{13}^2 \ud m_{23}^2 \ud\alpha \ud(\cos\beta) \ud\gamma$
which is consistent with Eq.(\ref{eq:FORA}).
    In the last row, the expressions of some intermediate variables defined to simplify the expressions of the three-momenta are given. 
    The $\lambda(x,y,z)$ called K$\ddot{a}$ll$\acute{e}$n triangle function is applied here as $\lambda(a^2,b^2,c^2)=(a^2-(b+c)^2)(a^2-(b-c)^2)$. The distribution under these IKVs is well known as Dalitz plot and its domain is given in the Kinematics Chapter of Review of Particle Physics (RPP) ~\cite{ParticleDataGroup:2020ssz}. 
    }
    \label{tab:3a}
\end{table}

\begin{table}[t]
    \centering
	\makegapedcells
        \begin{tabular}{|c|c|}
        \hline
        \textbf{IKVs} & 
        $m_{12},\Omega_3=(\cos\theta_3,\phi_3),\Omega_1^\star=(\cos\theta_1^\star,\phi_1^\star)$
        \\
        \hline
        \textbf{$A$} & 
         $ \frac{\abs{\boldsymbol{p_1^\star}}\abs{\boldsymbol{p_3}}}{\left(2\pi\right)^9 8m}       $
        \\
        \hline
        \textbf{$\boldsymbol{p}_{1,\,2}$} &
        $
        \begin{aligned}
            p_{1x}&=\normofvector{p_1^{\star}}\left(h\cos\phi_3-\sin\phi_3 k\right) + s_1\sin\theta_3 \cos\phi_3   
            \\
            p_{1y}&=\normofvector{p_1^{\star}}\left(h\sin\phi_3+\cos\phi_3k\right) + s_1\sin\theta_3 \sin\phi_3 \\
            p_{1z}&=-\normofvector{p_1^{\star}}\sin\theta_3\sin\theta_1^{\star}\cos\phi_1^{\star}    +s_1\cos\theta_3             \\
   p_{2x}&=\normofvector{p_1^{\star}}\left(-h\cos\phi_3+\sin\phi_3 k\right) +s_2\sin\theta_3\cos\phi_3     \\
   p_{2y}&=\normofvector{p_1^{\star}}\left(-h\sin\phi_3-\cos\phi_3k\right) +s_2 \sin\theta_3\sin\phi_3     \\
   p_{2z}&=\normofvector{p_1^{\star}}\sin\theta_3\sin\theta_1^{\star}\cos\phi_1^{\star} +s_2\cos\theta_3         \end{aligned}
        $
        \\
        \hline
        \textbf{}
        &
            $\begin{aligned}
                h&=\cos\theta_3\sin\theta_1^{\star}\cos\phi_1^{\star}, 
                \\
                k&=\sin\theta_1^{\star}\sin\phi_1^{\star}
                \\
                s_1&=-\gamma\beta\sqrt{m_1^2+\normofvector{p_1^{\star}}^2}+\gamma\normofvector{p_1^{\star}}\cos\theta_1^{\star}
                \\
                s_2&=-\gamma\beta\sqrt{m_2^2+\normofvector{p_1^{\star}}^2}-\gamma\normofvector{p_1^{\star}}\cos\theta_1^{\star}
                 \\
                \normofvector{p_1^{\star}}&=\frac{\lambda^\frac{1}{2}\left(m^2_{12},m^2_1,m^2_2\right)}{2m_{12}}
                \\
                \normofvector{p_3} &=\frac{\lambda^\frac{1}{2}\left(m^2,m^2_{12},m^2_3\right)}{2m}
                \\
                \gamma\beta &=\sqrt{\gamma^2 -1 }=\frac{\normofvector{p_3}}{m_{12}}
                \\

                        \end{aligned}$
        \\
        \hline
        \end{tabular}
\caption{This set of IKVs contains one invariant mass variable.
    $\Omega_3=(\cos\theta_3,\phi_3)$ and $\Omega_1^\star=(\cos\theta_1^\star,\phi_1^\star)$ are the solid angles of particle 3 in the rest frame of mass and particle 1 in the rest frame of composite particle-12, respectively.
    $A$ is the phase space factor and here 
    $\ud\Phi_3 = A\ud m_{12} \ud \Omega_3 \ud\Omega^\star_1$
    which is consistent with Eq.(\ref{eq:FORA}).
    In the last row, the expressions of some intermediate parameters defined to simplify the expressions of the three-momenta are given. The domain of the IKVs is $m_1 + m_2 \le m_{12} \le m-m_3$, $-1\le \cos\theta_3,\cos\theta_1^* \le 1$, and $0\le \phi_3,\phi_1^* \le 2\pi$.
     }
        \label{tab:3b}
\end{table}

\begin{table}[t]
    \centering
	\makegapedcells
        \begin{tabular}{|c|c|}
        \hline
        \textbf{IKVs} & 
        $\alpha,\cos\beta,\gamma,\theta_{12},\theta_{13}$
        \\
        \hline
        \textbf{$A$} & 
         $ \frac{\abs{\boldsymbol{p_1}}^2\abs{\boldsymbol{p_2}}^2\sin^2\theta_{12}}{8\left(2\pi\right)^9\left(E_2E_3\sin^2\theta_{13} + E_1E_3\sin^2(\theta_{12}+\theta_{13})+ E_1E_2 \sin^2\theta_{12}\right)}
        $
        \\
        \hline
        \textbf{$\boldsymbol{p_{1,\,2}}$} &
        $
        \begin{pmatrix*}
            p_{1x}\\
            p_{1y}\\
            p_{1z}\\
        \end{pmatrix*}
        = \mathcal{R}
        \begin{pmatrix*}
            0\\
            0\\
            \normofvector{p_1}\\
        \end{pmatrix*} $,
        $\begin{pmatrix*}
            p_{2x}\\
            p_{2y}\\
            p_{2z}\\
        \end{pmatrix*}
        =
        \mathcal{R}
        \begin{pmatrix*}
            \normofvector{p_2}\sin\theta_{12}\\
            0\\
            \normofvector{p_2}\cos\theta_{12}\\
        \end{pmatrix*}$
            \\
        \hline
        \textbf{}
        &
            $\begin{aligned}
                \normofvector{p_1} &= \frac{\sin\theta_{13}}{\sin\theta_{12}}\normofvector{p_3}
            \\
                \normofvector{p_2} &= -\frac{\sin\left(\theta_{12}+\theta_{13}\right)}{\sin\theta_{12}}\normofvector{p_3}
                \\
			\text{where}\;
                \sqrt{\normofvector{p_1}^2 + m_1^2} &+ \sqrt{\normofvector{p_2}^2+m_2^2} + \sqrt{\normofvector{p_3}^2 + m_3^2} = m
            \end{aligned}$
        \\
        \hline
        \end{tabular}
\caption{
    This set of IKVs does not contain any invariant mass variables.
    $A$ is the phase space factor and here 
    $\ud\Phi_3 = A \ud\alpha\ud(\cos\beta)\ud\gamma\ud\theta_{12}\ud\theta_{23}$
    which is consistent with Eq.(\ref{eq:FORA}).
    The domain of the IKVs is $0 \le \alpha\le2\pi$, $-1\le\cos\beta\le1$, $0\le\gamma\le2\pi$, $0\le\theta_{12}\le\pi$, and $\pi - \theta_{12}\le\theta_{13}\le\pi$
    }
     \label{tab:3c}
\end{table}

\section{The Phase space for Four-body final states}
\label{sec4}

{
\subsection{Invariant Mass Variables and Distribution Number(DN)}

For the four-body final states, there are six and four invariant masses variables for two ($m_{ij},i<j$) and three ($m_{ijk},i<j<k$) particles system, respectively.
However, only five of them are independent because of the following five equations,
\begin{align}
    \sum\limits_{j>i=1}^4 m_{ij}^2 &= m^2+2\sum\limits_{j=1}^4m_j^2,
    \label{eq:mij}
    \\
      m_{123}^2 &= m_{12}^2 + m_{23}^2 + m_{13}^2 - m_1^2 -m_2^2 - m_3^2
      \label{eq:m123}  
    \\
      m_{124}^2 &=  m_{12}^2 + m_{14}^2 + m_{24}^2 - m_1^2 -m_2^2 - m_4^2
      \label{eq:m124}  
    \\
      m_{134}^2 &=  m_{13}^2 + m_{14}^2 + m_{34}^2 - m_1^2 -m_3^2 - m_4^2
      \label{eq:m134}  
    \\
      m_{234}^2 &= m_{23}^2 + m_{24}^2 + m_{34}^2 - m_2^2 -m_3^2 - m_4^2
      \label{eq:m234}  
\end{align}
Therefore, up to 5 invariant masses can be chosen as IKVs.

In principle, there are $\sum_{i=1}^5 C^i_{10} = 462$ 
($C^a_b\equiv b!/(a!(b-a)!)$ is the combination number)
 different sets of the invariant mass.
However, lots of them are equivalent.
In order to classify all possible unique sets, a new concept of DN denoted by $(n;m;abcd)$ is introduced here.
Numbers in the bracket are of following meanings:
$n$ denotes the number of invariant masses and obviously satisfies the restriction $0\leq n \leq 5$; 
$abcd$ denotes the times that the particle index appears in the subscripts with $a\ge b\ge c\ge d$;
$m$ denotes the summation $a+b+c+d$.
For instance, for the set $\{m_{12}^2,m_{23}^2,m_{123}^2,\text{some angles}\}$,$n=3$, $m=7$ and $abcd=3220$.
Here $a=3$ for particle index 2 appears three times in the subscripts of three invariant mass variables, and $b=2$, $c=2$ and $d=0$ are for particle $1$, $3$ and $4$, respectively.   
Because of the restriction of $a\ge b\ge c\ge d$, cases that only different from the order of the particle indexes will correspond to the same DN.
For example, the sets $(m_{23}, m_{24}, m_{12}, m_{34}, m_{123})$ and $(m_{12}, m_{13}, m_{14}, m_{23}, m_{124})$ both correspond to the DN$=(5;11;4322)$, which means they can be converted into each other through changing the particle indexes from $(1234)$ to $(2341)$.
Therefore, the number of inequivalent sets of the invariant mass reduces to about 30 from 462.  
Typically, one DN may contains 2 different sets of IKVs.
Fortunately, it only happens with DN$=(4;9;3321)$ and $(3;7;2221)$.
Furthermore, some cases corresponding to different DNs are of the same kinematic structure because of Eqs.(\ref{eq:mij}-\ref{eq:m234}). 
For instance, if any $m_{ijk}$ is in the set containing five invariant masses, then it can be easily converted into the set containing five $m_{ij}$ whose DN$=(5;10;3322)$.
Table.(\ref{tab:conversion}) shows such conversion and the representative of each case is picked.
At last, there are 22 cases survived.

\begin{table}[H]
    \centering
	\makegapedcells
        \begin{tabular}{|c|c|c|}
        \hline
        \textbf{Others}
        & \textbf{Representative}
        & \textbf{Example} ($ij$ \textbf{is short for} $m_{ij}$)
        \\
        \hline
        $(5;m;abcd)$ & $(5;11;4322)$ & $12,13,14,23,24\rightarrow 12,13,14,23,124$
        \\
        \hline
        $(4;9;3222)$ & $(4;8;3221)$ & $12,13,34,124\rightarrow 12,13,23,34$
        \\
        \hline
        \multirow{2}{*}{$(4;9;3321)$} & $(4;10;4321)$ & $12,13,23,124\rightarrow 12,13,123,124$ 
        \\ \cline{2-3}
         & $(4;8;3221)$ & $12,13,24,123\rightarrow 12,13,23,24$ 
        \\ \hline
        $(4;10;3331)$ & $(4;8;3221)$ & $12,13,123,234\rightarrow 12,13,14,23$
        \\ \hline
        $(4;11;3332)$ & $(4;11;4322)$ & $12,124,134,234 \rightarrow 12,124,123,134$
        \\ \hline
        $(3;6;2220)$ & $(3;7;3220)$ & $12,14,24\rightarrow 12,14,124$
        \\ \hline
        $(3;7;2221)$ & $(3;6;3111)$ & $12,13,234\rightarrow 12,13,14$
        \\ \hline
        $(3;8;2222)$ & $(3;7;2222)$ & $12,134,234 \rightarrow 12,34,234$
        \\  \hline
        \end{tabular}
        \caption{Cases in the ``others'' column can be easily converted into the case in the ``Representative'' column. 
        There are two distinct cases with DN$=(4;9;3321)$ as well as $(3;7;2221)$, 
        where the particle corresponding to $d=1$ can appear in $m_{ij}$ or $m_{ijk}$. 
        }
\label{tab:conversion}
\end{table}

\subsection{Simplification of expressions of three-momenta of final particles}
\label{sub2sec2}

In our notation, if $\abs{\boldsymbol{p_i}}$ and $\theta_{ij}$ for each particle are all known, general expressions of components of three-momenta in terms of Euler angles can be calculated as follows
\begin{align}
    \begin{pmatrix*}
        p_{1x}\\
        p_{1y}\\
        p_{1z}\\
    \end{pmatrix*}
    &= \mathcal{R}
    \begin{pmatrix*}
        0\\
        0\\
        \normofvector{p_1}\\
    \end{pmatrix*},
    \label{eq:p1xR}\\
    \begin{pmatrix*}
        p_{2x}\\
        p_{2y}\\
        p_{2z}\\
    \end{pmatrix*}
    &=
    \mathcal{R}
    \begin{pmatrix*}
        \normofvector{p_2}\sin\theta_{12}\\
        0\\
        \normofvector{p_2}\cos\theta_{12}\\
    \end{pmatrix*},
    \label{eq:p2xR}\\
    \begin{pmatrix*}
        p_{3x}\\
        p_{3y}\\
        p_{3z}\\
    \end{pmatrix*}
    &=
    \mathcal{R}
    \begin{pmatrix*}
        \normofvector{p_3}\frac{\cos\theta_{23}-\cos\theta_{13}\cos\theta_{12}}{\sin\theta_{12}}\\
        \pm\normofvector{p_3}\frac{1}{A_g\sin\theta_{12}}\\
        \normofvector{p_3}\cos\theta_{13}\\
    \end{pmatrix*},
   \label{eq:p3xR}\\
    \begin{pmatrix*}
        p_{4x}\\
        p_{4y}\\
        p_{4z}\\
    \end{pmatrix*}
    &=
    \mathcal{R}
    \begin{pmatrix*}
        -p_{2x}-p_{3x}\\
        -p_{3y}\\
        -p_{1z}-p_{2z}-p_{3z}\\
    \end{pmatrix*},
\label{eq:p4xR}
\end{align}
where $\mathcal{R}$ is defined in Eq.(\ref{eq:R}), and $A_g$ is defined as,
\begin{align}
A_g = \frac{1}
{\sqrt{1+2\cos\theta_{13}\cos\theta_{12}\cos\theta_{23}
       -\cos^2\theta_{12}-\cos^2\theta_{13}-\cos^2\theta_{23}
       }
}.
\label{eq:Ag}
\end{align} 
It is obvious that we just need six variables, including $\abs{\boldsymbol{p_1}}$, $\abs{\boldsymbol{p_2}}$, $\abs{\boldsymbol{p_3}}$, $\theta_{12}$, $\theta_{23}$  and $\theta_{13}$, to compute all three-momenta of final states.
Note that there are two choices with different sign for the $p_{3y}$.
It corresponds to the two allowed patterns if just $\theta_{ij}$ and $|\boldsymbol{p_i}|$ are fixed, as shown in Fig.\ref{fg:mirror} where Euler angles have been chosen as $(\alpha,\beta,\gamma)=(0,0,0)$, 
Actually, two configurations in the Fig.(\ref{fg:mirror}) are indistinguishable with respect to the IKVs we have chosen.
To avoid this arbitrariness, some more variables denoting the sign of $(\boldsymbol{p}_1\times\boldsymbol{p}_2)\cdot\boldsymbol{p}_3$ are needed. 
However, it is unnecessary because the phase space factors $A$ for these two configurations are exactly the same.
Thus, one set of IKVs will give at least two sets of the three-momenta of final states, and then the amplitudes of these two three-momenta could be different.
We should re-write Eq.(\ref{eq:FORA}) as follows,
\begin{align}
    |\mathcal{M}|^2\ud\Phi_4  = A(|\mathcal{M}^-|^2+|\mathcal{M}^+|^2)\, \ud m_a \ud m_b \cdots \ud \alpha_1 \ud\alpha_2 \cdots, 
    \label{eq:FORA2}
\end{align} 
where $\mathcal{M}^\pm$ are for the amplitudes with different sign of $p_{3y}$.

\begin{figure}[H]
    \centering
    \includegraphics[scale=1]{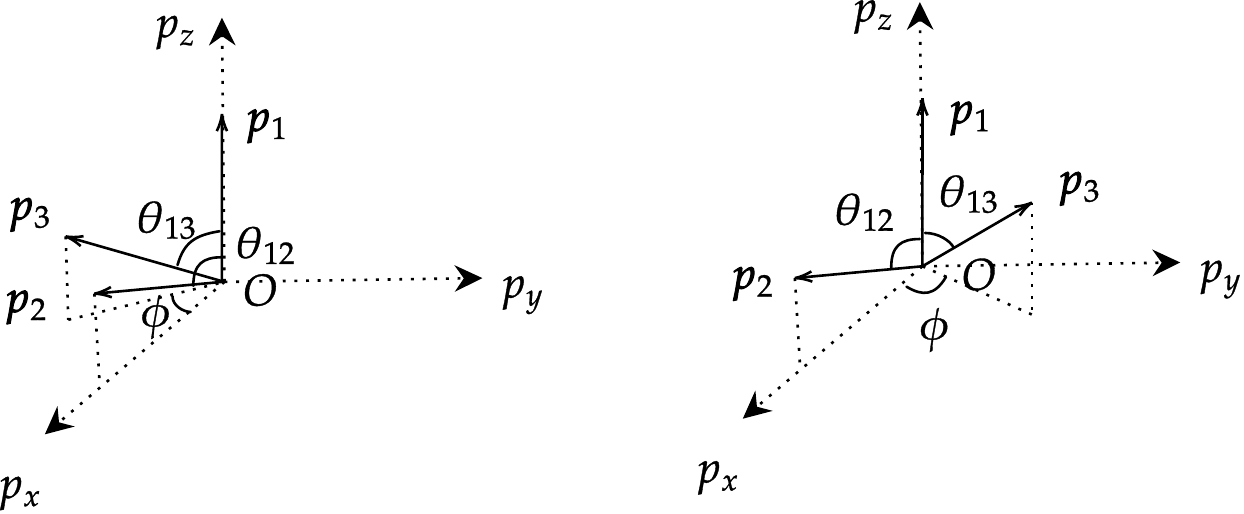}
\caption{
Both patterns are allowed if just $\theta_{ij}$ and $|\boldsymbol{p_i}|$ are fixed. 
For the left plot, $\left(\boldsymbol{p_1}\times\boldsymbol{p_2}\right)\cdot\boldsymbol{p_3}$ is negative while for the right one it's positive.
Euler angles here have been chosen as $(\alpha,\beta,\gamma)=(0,0,0)$.
$\boldsymbol{p_4}=-(\boldsymbol{p_1}+\boldsymbol{p_2}+\boldsymbol{p_3})$ is not shown here. 
Two configurations can be converted by the mirror reflection with respect to the $\boldsymbol{p_1} O \boldsymbol{p_2}$ plane.}
\label{fg:mirror}
\end{figure}

Furthermore, the six variables $\abs{\boldsymbol{p_1}}$, $\abs{\boldsymbol{p_2}}$, $\abs{\boldsymbol{p_3}}$, $\theta_{12}$, $\theta_{23}$  and $\theta_{13}$ can be computed by the three energies $E_{1,\,2,\,3}$ and three invariant masses $m_{12},\,m_{13},\,m_{23}$ as follows, 
\begin{align}
\normofvector{p_{i}}&=\sqrt{E_i^2-m_i^2}\,,\label{eq:e}\\
\cos\theta_{ij} &= \frac{2E_i E_j + m_i^2 + m_j^2 - m_{ij}^2}{2\normofvector{p_i}\normofvector{p_j}}.
\label{eq:cos}
\end{align} 
Therefore, it is found that if three energies $E_{1,\,2,\,3}$ and three invariant masses $m_{12},\,m_{13},\,m_{23}$ are given, all components of the three-momenta of the final states can be computed.
In the last subsection, the relationship between these six physical quantities and IKVs will be given.

By the way, for DN=$(3;6;3111)$ and $(2;5;2111)$, there are two possible solutions for the $E_{2}$.
It is easy to understand that in these two sets, $m_{12}$, $\cos\theta_{12}$ and $E_1$ can be fixed by IKVs.
Then, $E_2$ can be solved from Eqs.(\ref{eq:e}) and (\ref{eq:cos}), 
\begin{align}
2\sqrt{E_1^2-m_1^2}\sqrt{E_2^2-m_2^2}\cos \theta_{12} = 2E_1 E_2 + m_1^2 + m_2^2 - m_{12}^2.
\label{eq:forE2}
\end{align} 
Obviously, there are two possible solutions for $E_2$, labeled as $E^+_2$ and $E^-_2$ which can be found explicit expressions in Tabels.(\ref{tab:3111}) and (\ref{tab:2111}).
At that time, the phase space factor should be re-defined as follows,  
\begin{align}
    |\mathcal{M}|^2\ud\Phi_4  = 
    \left[A(E^+_2)\left(|\mathcal{M}^-(E^+_2)|^2+|\mathcal{M}^+(E^+_2)|^2\right) + 
     A(E^-_2)\left(|\mathcal{M}^-(E^-_2)|^2+|\mathcal{M}^+(E^-_2)|^2\right)\right]
    \, \ud m_a \ud m_b \cdots \ud \alpha_1 \ud\alpha_2 \cdots.
    \label{eq:FORA3}
\end{align} 
%


\subsection{Domain of integration}

To complete the integration of the phase space, the domains of IKVs are also needed.
The domain of Euler angles are trivial, that's to say, $0\leq\alpha\leq2\pi$, $0\leq\beta\leq\pi$, and $0\leq\gamma\leq2\pi$.
As for the other IKVs, things become much more complicated. 
Among the cases given in this paper, some cases with four or five invariant masses chosen as IKVs are relatively simple since the calculation can be reduced to the three-body final states. 
Two examples are given here and the similar discussion can be found in Ref.~\cite{PhysRev.185.1865}. For the set $\left(m_{124}^2,m_{12}^2,m_{14}^2,m_{13}^2,m_{23}^2\right)$, one can complete the integration as follows,
\begin{align}
\int\limits_{(m_1+m_2+m_4)^2}^{(m-m_3)^2} \ud m_{124}^2
\int\limits_{(m_1+m_2)^2}^{(m_{124}-m_4)^2} \ud m_{12}^2
\int\limits_{C_1^+}^{C_1^-} \ud m_{14}^2
\int\limits_{C_2^+}^{C_2^-} \ud m_{13}^2
\int\limits_{C_3^+}^{C_3^-} \ud m_{23}^2,
\end{align}
where
\begin{align}
C_1^\pm &= (E_1^*+E_4^*)^2 - (\sqrt{E_1^{*2}-m_1^2} \pm \sqrt{E_4^{*2} - m_4^2},\\
C_2^\pm &= (\tilde{E}_1^* + \tilde{E}_3^*)^2 - (\sqrt{\tilde{E}_1^{*2}-m_1^2} \pm \sqrt{\tilde{E}_3^{*2} - m_3^2} ),\\
C_3^\pm &=(\hat{E}_2^* + \hat{E}_3^*)^2 - (\sqrt{\hat{E}_2^{*2}-m_2^2} \pm \sqrt{\hat{E}_3^{*2} - m_3^2} ),
\end{align}
with
\begin{align}
E_1^* &= (m_{12}^2 - m_{1}^2 + m_2^2 )/(2m_{12}),\\
E_4^* &= (m_{124}^2 - m_{12}^2 -m_4^2)/(2m_{12}),\\
\tilde{E}_1^* &= (m_{124}^2 - m_{24}^2 + m_{3}^2)/(2m_{124}),\\
\tilde{E}_3^* &= (m^2 - m_{124}^2 - m_3^2)/(2m_{124}),\\
\hat{E}_2 &= (m_{24}^2 - m_4^2 + m_2^2)/(2m_{24}),\\
\hat{E}_3 &= (m_{234}^2 - m_{24}^2 -m_3^2)/(2m_{24}),\\
m_{24}^2 &= m_{124}^2 + m_1^2 + m_2^2 + m_4^2 - m_{12}^2 - m_{14}^2, \\
m_{234}^2 &= m^2 + m_1^2 + \sum\limits_{i=1}^4 m_i^2 - m_{12}^2 - m_{13}^2 -m_{14}^2.
\end{align}

As another example, for the set $\left(m_{12}^2,m_{23}^2,m_{24}^2,m_{234}^2,\cos\theta_3^*\right)$, one can complete the integration as follows,
\begin{align}
\int\limits_{-1}^1 \ud\cos\theta_3^*
\int\limits_{(m_2 + m_3 + m_4)^2}^{(m-m_1)^2} \ud m_{234}^2
\int\limits_{(m_2+m_3)^2}^{(m_{234}-m_4)^2} \ud m_{23}^2
\int\limits_{C_1^+}^{C_1^-} \ud m_{24}^2
\int\limits_{C_2^+}^{C_2^-} \ud m_{12}^2,
\end{align}
where
\begin{align}
C_1^\pm &= (E_2^*+E_4^*)^2 - (\sqrt{E_2^{*2}-m_2^2} \pm \sqrt{E_4^{*2} - m_4^2} ),\\
C_2^\pm &= (\tilde{E}_2^* + \tilde{E}_1^*)^2 - (\sqrt{\tilde{E}_2^{*2}-m_2^2} \pm \sqrt{\tilde{E}_1^{*2} - m_1^2} ),
\end{align}
with
\begin{align}
{E}_2^* &= (m_{23}^2 - m_3^2 + m_2^2)/(2m_{23}),\\
E_4^* &= (m_{234}^2 - m_{23}^2 - m_4^2) / (2m_{23}) ,\\
\tilde{E}_2^* &= (m_{234}^2 - m_{34}^2 + m_2^2)/(2m_{234}),\\
\tilde{E}_1^* &= (m^2 - m_{234}^2 - m_1^2)/(2m_{234}),\\
m_{34}^2 &= m_{234}^2 + m_2^2 + m_3^2 + m_4^2 - m_{23}^2 -m_{24}^2.
\end{align}
However, it is found that the explicit domain functions of IKVs here would become much complicated if the order of the IKVs changed. 
For example, it is really not easy to obtain the explicit domain functions of $m^2_{23}$
if only $m^2_{124}$ is fixed.
For the other sets of the IKVs, especially which can not be reduced into three-body final states, the domain can not be obtained without tedious calculation.
Fortunately, in the numerical calculation, we do not need such explicit domain functions of each IKVs and here another numerical method is introduced as follows.
It is more practical to do the integration under the following restrictions that completely determine the boundary of the phase space.
The rough ranges of the angles except Euler angles are,
\begin{align}
0\leq\theta_{ij}\leq\pi,\label{eq:angular1}\\
0\leq\theta_i^\star\leq\pi.\label{eq:angular2}
\end{align}
The invariant mass variables are supposed to satisfy the following restrictions at least,
\begin{gather}
\left(m_i+m_j\right)^2 \leq m_{ij}^2\leq \left(m-\sum\limits_{k\neq i,j}m_k\right)^2,\label{eq:invmass1}\\
\left(m_i+m_k+m_l\right)^2 \leq\left(m_j+m_{kl}\right)^2\leq m^2_{jkl}\leq\left(m-\sum_{i\neq j,k,l} m_i\right)^2,\label{eq:invmass2}
\end{gather}
where $m_{kl}\geq m_{jk} \geq m_{jl}$ is assumed.
Then, to obtain the exact range of variables, we are supposed to check whether the values of some physical quantities expressed by the IKVs are physical or not step by step.
Firstly, the energy and the mass of any particle should satisfy, 
\begin{equation}\label{eq:eimi}
E_i\geq m_i .
\end{equation}
Secondly, another natural restriction on the angle between two final particles $\theta_{ij}$ is, 
\begin{equation}\label{eq:cosij}
|\cos\theta_{ij}|= \frac{\left|2E_i E_j + m_i^2 + m_j^2 - m_{ij}^2\right|}{2\normofvector{p_i}\normofvector{p_j}}\leq 1.
\end{equation}
Thirdly, the factor $A_g$ in the expression of $p_{3y}$ in Eq.(\ref{eq:p3xR}) should satisfy the following restriction to ensure the reality of $p_{3y}$, 
\begin{align}
1+2 \cos \theta_{13} \cos \theta_{12} \cos \theta_{23}-\cos ^{2} \theta_{12}-\cos ^{2} \theta_{13}-\cos ^{2} \theta_{23}\geq 0	.
\label{eq:Agcon}
\end{align}
For all IKVs listed in this paper for the four-body final states, the restrictions above are enough to control the integration ranges of IKVs.
In the numerical calculation, one can first sampling within the rough region of Eqs.(\ref{eq:angular1} - \ref{eq:invmass2}) for the IKVs and then only sum the contributions of the samples which satisfy the physical conditions of Eqs.(\ref{eq:eimi}-\ref{eq:Agcon}).
Typically, it is worthy to note that the variables in the restrictions can be easily calculated from the three momenta of the final states which are explicitly expressed by IKVs.

\subsection{Formulae}

In this section, the formulae of all cases of IKVs of four-body final states are listed in Tables.(\ref{tab:4322}-\ref{tab:1110}). 
In each table, the expressions of three energies $E_{1,\,2,\,3}$ and three invariant masses $m_{12,\,13,\,23}$ are shown as discussed before.
Furthermore, some other intermediate variables which are defined to simplify the expressions of $E_{1,\,2,\,3}$ and $m_{12,\,13,\,23}$ are given in the last rows of the corresponding tables. Euler angles is not included in IKVs since they are supposed to appear in all cases.
For some cases with DN$=(2;m;abcd)$, an equation is given in the last rows of the corresponding table. Though the analytical solution exists, the expression is so complicated that it will not be given. Besides, for those cases, $\theta_{i(jk)}$ denotes the angle between $\boldsymbol{p_i}$ and $\boldsymbol{p_j} + \boldsymbol{p_k}$ and $E_{ij}$ is short for $E_i + E_j$.

In these tables, the phase space factors and the three-momenta of the final states are shown explicitly.
Then, once we have the formulae of amplitudes, the differential cross-section and differential decay can be calculated by using Eq.(\ref{eq:dsigma}) and Eq.(\ref{eq:dGamma}), respectively.
%

\begin{table}[H]
    \centering
	\makegapedcells
        \begin{tabular}{|c|c|}
        \hline
        $\begin{gathered}
            \textbf{IKVs}
            \\
            \textbf{DN=(5;11;4322)}
        \end{gathered} $
        & 
        $m_{12}^2$,$m_{13}^2$,$m_{14}^2$,$m_{23}^2$,$m_{124}^2$
        \\
        \hline
        \textbf{$A$} & 
    $   \frac{A_g}{\left(2\pi\right)^{12}2^9m^3\abs{\boldsymbol{p_1}}\abs{\boldsymbol{p_2}}\abs{\boldsymbol{p_3}}}
    $
        \\
        \hline
        $\begin{gathered}
             E_{1,\,2,\,3}
            \\
            \text{and}
            \\
             m_{12,\,13,\,23}    
        \end{gathered} $
 &
        $\begin{aligned}
            E_1 &= \frac{1}{2m}\left(m_{12}^2 + m_{13}^2 + m_{14}^2 - \sum\limits_{i=1}^4 m_i^2\right)
            \\
            E_2 &= \frac{1}{2m}\left(m_{23}^2 + m_{124}^2 - m_{14}^2 - m_3^2 \right)
            \\
            E_3 &= \frac{1}{2m}\left(m^2 - m_{3}^2 - m_{124}^2 \right)
            \\     
		&\text{$m_{12,\,13,\,23}$ are IKVs directly}   
        \end{aligned}
        $
        \\
        \hline
        \end{tabular}
\caption{
This set of IKVs contains five invariant mass variables and the corresponding DN is $(5;11;4322)$. 
$A$ is the phase space factor and here $\ud\Phi_4 = A \ud\alpha\ud(\cos\beta)\ud\gamma\ud m^2_{12}\ud m^2_{13} \ud m^2_{14}\ud m^2_{23}\ud m_{124}^2$ which is consistent with Eq.(\ref{eq:FORA2}).
}
\label{tab:4322}
\end{table}    
}

\begin{table}[H]
    \centering
	\makegapedcells
        \begin{tabular}{|c|c|}
        \hline
		$\begin{gathered}
			\textbf{IKVs}
			\\
			\textbf{DN=(4;8;3221)}
		\end{gathered}$
		&
		$m_{12}^2$,$m_{13}^2$,$m_{14}^2$,$m_{34}^2$,$\cos\theta_3^\star$
        \\
        \hline
   		$A$ & 
    $  \frac{A_g\gammaL\betaL\abs{\boldsymbol{p_3^\star}}}{\left(2\pi\right)^{12}2^8m^2\abs{\boldsymbol{p_1}}\abs{\boldsymbol{p_2}}\abs{\boldsymbol{p_3}}}
    $
        \\
        \hline
        $\begin{gathered}
             E_{1,\,2,\,3}
            \\
            \text{and}
            \\
             m_{12,\,13,\,23}    
        \end{gathered} $ &
        $\begin{aligned}
            E_1 &= \frac{1}{2m}\left(m_{12}^2 + m_{13}^2 + m_{14}^2 - \sum\limits_{i=1}^4 m_i^2\right)
            \\
            E_2 &= \frac{1}{2m}\left(m^2 - m_{13}^2 - m_{14}^2 - m_{34}^2 + \sum\limits_{i=1}^4 m_i^2\right)
            \\
            E_3 &= \gammaL\sqrt{\normofvector{p_3^\star}^2+m_3^2} - \gammaL\betaL\normofvector{p_3^\star}\cos\theta_3^\star
            \\
            m_{23}^2 &= 2mE_3 - \sum\limits_{i=1}^4 m_i^2 - m_{13}^2 - m_{34}^2
			\\
		&\text{$m_{12,\,23} $ are IKVs directly}
        \end{aligned}
        $
        \\
        \hline
        {} &
        $\begin{aligned}
           \gammaL\betaL &= \sqrt{\gammaL^2 -1 } =  \frac{1}{2mm_{34}} \lambda^\frac{1}{2} \left(m^2,m^2_{12},m^2_{34}\right)
			\\
            \normofvector{p_3^\star} &= \frac{1}{2m_{34}} \lambda^\frac{1}{2} \left(m^2_{34},m^2_3,m^2_4\right)        
        \end{aligned}
        $
        \\
        \hline
        \end{tabular}
    \caption{This set of IKVs contains four invariant mass variables and the corresponding DN is $(4;8;3221)$. 
$A$ is the phase space factor and here $\ud\Phi_4 = A \ud\alpha\ud(\cos\beta)\ud\gamma\ud m^2_{12}\ud m^2_{13} \ud m^2_{14}\ud m^2_{34}\ud (\cos\theta_3^\star)$ which is consistent with Eq.(\ref{eq:FORA2}).
Quantities with superscripts $\star$ are defined in the rest frame of the composite particle-34.  
    }
\end{table}


\begin{table}[H]
    \centering
	\makegapedcells
        \begin{tabular}{|c|c|}
		\hline
		$\begin{gathered}
			\textbf{IKVs}
			\\
			\textbf{DN=(4;8;2222)}
		\end{gathered}$
		&
		$m_{12}^2$,$m_{13}^2$,$m_{24}^2$,$m_{34}^2$,$\cos\theta_1^\star$
        \\
        \hline
        ${A}$& 
    $      \frac{A_g\gammaL\betaL\abs{\boldsymbol{p_1^\star}}}{\left(2\pi\right)^{12}2^8m^2\abs{\boldsymbol{p_1}}\abs{\boldsymbol{p_2}}\abs{\boldsymbol{p_3}}}
    $
        \\
        \hline
        $\begin{gathered}
             E_{1,\,2,\,3}
            \\
            \text{and}
            \\
             m_{12,\,13,\,23}    
        \end{gathered} $ &
        $\begin{aligned}
            E_1 &=  \gammaL\sqrt{\normofvector{p_1^\star}^2+m_1^2} - \gammaL\betaL\normofvector{p_1^\star}\cos\theta_1^\star
            \\
            E_2 &= \gammaL\sqrt{\normofvector{p_1^\star}^2+m_2^2} + \gammaL\betaL\normofvector{p_1^\star}\cos\theta_1^\star
            \\
            E_3 &= \sqrt{\frac{\lambda\left(m^2,m^2_{13},m^2_{24}\right)}{4m^2}+m_{13}^2} - E_1
            \\
            m_{23}^2 &= 2mE_2 - \sum\limits_{i=1}^4 m_i^2 - m_{12}^2 - m_{24}^2
			\\
			&\text{$m_{12,23}$ are IKVs directly}
                \end{aligned}
        $
        \\
        \hline
        { } &
        $\begin{aligned}
            \gammaL\betaL &= \sqrt{\gammaL^2-1} =\frac{1}{2mm_{12}} \lambda^\frac{1}{2} \left(m^2,m^2_{12},m^2_{34}\right)
            \\
       |\boldsymbol{p_1^\star}| & = \frac{1}{2m_{12}}\lambda^{\frac{1}{2}}\left(m^2_{12},m_1^2,m_2^2 \right)
        \end{aligned}
        $
        \\
        \hline
        \end{tabular}
    \caption{This set of IKVs contains four invariant mass variables and the corresponding DN is $(4;8;2222)$. 
$A$ is the phase space factor and here $\ud\Phi_4 = A \ud\alpha\ud(\cos\beta)\ud\gamma\ud m^2_{12}\ud m^2_{13} \ud m^2_{24}\ud m^2_{34}\ud (\cos\theta_1^\star)$ which is consistent with Eq.(\ref{eq:FORA2}).
Quantities with superscripts $\star$ are defined in the rest frame of the composite particle-12.
}
\end{table}

\begin{table}[H]
    \centering
	\makegapedcells
        \begin{tabular}{|c|c|}
        \hline
		$\begin{gathered}
			\textbf{IKVs}
			\\
			\textbf{DN=(4;9;4221)}
		\end{gathered}$
		&  $m_{12}^2$,$m_{23}^2$,$m_{24}^2$,$m_{234}^2$,$\cos\theta_3^\star$
        \\
        \hline
        $A$ & 
    $    \frac{A_g\gammaL\betaL\abs{\boldsymbol{p_3^\star}}}{\left(2\pi\right)^{12}2^8m^2\abs{\boldsymbol{p_1}}\abs{\boldsymbol{p_2}}\abs{\boldsymbol{p_3}}}
    $
        \\
        \hline
         $\begin{gathered}
             E_{1,\,2,\,3}
            \\
            \text{and}
            \\
             m_{12,\,13,\,23}    
        \end{gathered} $ &
        $\begin{aligned}
            E_1 &= \frac{1}{2m} \left(m^2 + m_1^2 - m_{234}^2\right)
            \\
            E_2 &= \frac{1}{2m}\left(m_{12}^2 + m_{23}^2 + m_{24}^2 - \sum\limits_{i=1}^4 m_i^2\right)
            \\
            E_3 &= \gammaL\sqrt{\normofvector{p_3^\star}^2+m_3^2} - \gammaL\betaL\normofvector{p_3^\star}\cos\theta_3^\star
            \\   
            m_{13}^2 &= 2m\left(E_1+E_2+E_3\right) -m^2 + \sum\limits_{i=1}^4 m_i^2 - m_{12}^2 - m_{23}^2 
            \\
			&\text{$m_{12,23}$ are IKVs directly } 
        \end{aligned}
        $
        \\
        \hline
       	{} &
        $\begin{aligned}
            \gammaL\betaL &= \sqrt{\gammaL^2 - 1} = \frac{1}{2mm_{234}}\lambda^\frac{1}{2}\left(m^2,m^2_1,m^2_{234}\right)
            \\
            \normofvector{p_3^\star} &= \frac{1}{2m_{234}}\lambda^\frac{1}{2}\left(m^2_{234},m^2_{24},m^2_3\right)
        \end{aligned}
        $
        \\
        \hline
        \end{tabular}
    \caption{This set of IKVs contains four invariant mass variables and the corresponding DN is $(4;9;4221)$. 
$A$ is the phase space factor and here $\ud\Phi_4 = A \ud\alpha\ud(\cos\beta)\ud\gamma\ud m^2_{12}\ud m^2_{23} \ud m^2_{24}\ud m^2_{234}\ud (\cos\theta_3^\star)$ which is consistent with Eq.(\ref{eq:FORA2}).
Quantities with superscripts $\star$ are defined in the rest frame of the composite particle-234.}
\end{table}

\begin{table}[H]
    \centering
	\makegapedcells
        \begin{tabular}{|c|c|}
        \hline
		$\begin{gathered}
			\textbf{IKVs}
			\\
			\textbf{DN=(4;10;4222)}
		\end{gathered}$
		&  $m_{12}^2$,$m_{13}^2$,$m_{124}^2$,$m_{134}^2$,$\cos\theta_4^\star$
        \\
        \hline
        $A$ & 
    $  \frac{A_g\gammaL\betaL\abs{\boldsymbol{p_4^\star}}}{\left(2\pi\right)^{12}2^8m^2\abs{\boldsymbol{p_1}}\abs{\boldsymbol{p_2}}\abs{\boldsymbol{p_3}}}
    $
        \\
        \hline
         $\begin{gathered}
             E_{1,\,2,\,3}
            \\
            \text{and}
            \\
             m_{12,\,13,\,23}    
        \end{gathered} $  &
        $\begin{aligned}
            E_1 &= m - \left(E_2 + E_3 + E_4\right)
            \\
            E_2 &= \frac{1}{2m}\left(m^2 + m_2^2 - m_{134}^2\right)
            \\
            E_3 &= \frac{1}{2m}\left(m^2 + m_3^2 - m_{124}^2\right)
            \\
            m_{23}^2 &= m^2 + \sum\limits_{i=1}^4 m_i^2 - m_{12}^2 - m_{13}^2 - 2mE_4
			\\
			&\text{$m_{12,13}$ are IKVs directly}
        \end{aligned}
        $
        \\
        \hline
       {} &
        $\begin{aligned}
            \gammaL\betaL &=\sqrt{\gammaL^2 -1 } = \frac{1}{2mm_{124}} \lambda^\frac{1}{2} \left(m^2,m^2_{3},m^2_{124}\right)
            \\
            \normofvector{p_4^\star} &= \frac{1}{2m_{124}} \lambda^\frac{1}{2} \left(m^2_{124},m^2_{12},m^2_{4}\right)
			\\
            E_4 &= \gammaL\sqrt{\normofvector{p_4^\star}^2+m_4^2} - \gammaL\betaL\normofvector{p_4^\star}\cos\theta_4^\star
            \end{aligned}
        $
        \\
        \hline
        \end{tabular}
   \caption{This set of IKVs contains four invariant mass variables and the corresponding DN is $(4;10;4222)$. 
$A$ is the phase space factor and here it is $\ud\Phi_4 = A\ud\alpha\ud(\cos\beta)\ud\gamma\ud m^2_{12}\ud m^2_{13} \ud m^2_{124}\ud m^2_{134}\ud (\cos\theta_4^\star)$ which is consistent with Eq.(\ref{eq:FORA2}).
Quantities with superscripts $\star$ are defined in the rest frame of the composite particle-124.}
\end{table}


\begin{table}[H]
    \centering
	\makegapedcells
        \begin{tabular}{|c|c|}
        \hline
		$\begin{gathered}
			\textbf{IKVs}
			\\
			\textbf{DN=(4;10;4321)}
		\end{gathered}$
		&  $m_{12}^2$,$m_{13}^2$,$m_{123}^2$,$m_{124}^2$,$\cos\theta_2^\star$
		\\
        \hline
        $A$ & 
    $ \frac{A_g\gammaL\betaL\abs{\boldsymbol{p_2^\star}}}{\left(2\pi\right)^{12}2^8m^2\abs{\boldsymbol{p_1}}\abs{\boldsymbol{p_2}}\abs{\boldsymbol{p_3}}}
    $
        \\
        \hline
        $\begin{gathered}
             E_{1,\,2,\,3}
            \\
            \text{and}
            \\
             m_{12,\,13,\,23}    
        \end{gathered} $ &
        $\begin{aligned}
            E_1 &= \frac{1}{2m}\left(m_{123}^2 + m_{124}^2 -m_3^2 - m_4^2 - 2mE_2\right)
            \\ 
            E_2 &= \gammaL\sqrt{\normofvector{p_2^\star}^2+m_2^2} - \gammaL\betaL\normofvector{p_2^\star}\cos\theta_2^\star
            \\
            E_3 &= \frac{1}{2m} \left(m^2 + m_3^2 - m_{124}^2\right)
            \\
            m_{23}^2 &= m_1^2 + m_2^2 + m_3^2 + m_{123}^2 -m_{12}^2 - m_{13}^2
			\\
			&\text{$m_{12,13}$ are IKVs directly}
                \end{aligned}
        $
        \\
        \hline
        {} &
        $\begin{aligned}
            \gammaL\betaL &= \sqrt{\gammaL^2 - 1}= \frac{1}{2mm_{123}} \lambda^\frac{1}{2} \left(m^2,m^2_{123},m^2_{4}\right)
            \\
            \normofvector{p_2^\star}& = \frac{1}{2m_{123}}\lambda^\frac{1}{2}\left(m^2_{123},m^2_{13},m^2_2\right)
        \end{aligned}
        $
        \\
        \hline
        \end{tabular}
    \caption{This set of IKVs contains four invariant mass variables and the corresponding DN is $(4;10;4321)$. 
$A$ is the phase space factor and here $\ud\Phi_4 = A \ud\alpha\ud(\cos\beta)\ud\gamma\ud m^2_{12}\ud m^2_{13} \ud m^2_{123}\ud m^2_{124}\ud (\cos\theta_2^\star)$ which is consistent with Eq.(\ref{eq:FORA2}).
Quantities with superscripts $\star$ are defined in the rest frame of the composite particle-123.}
\end{table}


\begin{table}[H]
    \centering
	\makegapedcells
        \begin{tabular}{|c|c|}
        \hline
		$\begin{gathered}
			\textbf{IKVs}
			\\
			\textbf{DN=(4;10;3322)}
		\end{gathered}$
		&  $m_{12}^2$,$m_{34}^2$,$m_{234}^2$,$m_{124}^2$,$\cos\theta_{13}$
        \\
        \hline
        $A$ & 
    $   \frac{A_g}{\left(2\pi\right)^{12}2^8m^3\abs{\boldsymbol{p_2}}}
    $
        \\
        \hline
        $\begin{gathered}
             E_{1,\,2,\,3}
            \\
            \text{and}
            \\
             m_{12,\,13,\,23}    
        \end{gathered} $ &
        $\begin{aligned}
            E_1 &= \frac{1}{2m}\left(m^2 + m_1^2 - m_{234}^2\right)
            \\
            E_2 &= \sqrt{\frac{\lambda\left(m^2,m^2_{12},m^2_{34}\right)}{4m^2} + m_{12}^2} - E_1
            \\
            E_3 &= \frac{1}{2m}\left(m^2 + m_3^2 - m_{124}^2\right)
            \\
            m_{13}^2 &= m_1^2 + m_3^2 + 2E_1E_3 - 2\normofvector{p_1}\normofvector{p_3}\cos\theta_{13}
            \\
            m_{23}^2 &= m_1^2 + m_2^2 + m_4^2 - m_{13}^2 - m_{34}^2 + m_{124}^2
			\\
		&\text{$m_{12}$ is IKV directly}
            \end{aligned}
        $
        \\
        \hline
        \end{tabular}
\caption{This set of IKVs contains four invariant mass variables and the corresponding DN is $(4;10;3322)$. 
$A$ is the phase space factor and here $\ud\Phi_4 = A \ud\alpha\ud(\cos\beta)\ud\gamma\ud m^2_{12}\ud m^2_{34} \ud m^2_{234}\ud m^2_{124}\ud (\cos\theta_{13})$ which is consistent with Eq.(\ref{eq:FORA2}).}
\end{table}


\begin{table}[H]
    \centering
	\makegapedcells
        \begin{tabular}{|c|c|}
        \hline
		$\begin{gathered}
			\textbf{IKVs}
			\\
			\textbf{DN=(4;11;4322)}
		\end{gathered}$
		&  $m_{12}^2$,$m_{124}^2$,$m_{123}^2$,$m_{134}^2$,$\cos\theta_{13}$
                \\
        \hline
        $A$ & 
    $  \frac{A_g}{\left(2\pi\right)^{12}2^8m^3\abs{\boldsymbol{p_2}}}
    $
        \\
        \hline
         $\begin{gathered}
             E_{1,\,2,\,3}
            \\
            \text{and}
            \\
             m_{12,\,13,\,23}    
        \end{gathered} $ &
        $\begin{aligned}
            E_1 &= \frac{1}{2m} \left(m_{123}^2 + m_{124}^2 + m_{134}^2 -m_2^2 - m_3^2 - m_4^2 -m^2\right)
            \\
            E_2 &= \frac{1}{2m} \left(m^2 + m_2^2 - m_{134}^2\right)
            \\
           	E_3 &= \frac{1}{2m} \left(m^2 + m_3^2 - m_{124}^2\right)
            \\
            m_{13}^2 &= m_1^2 + m_3^2 + 2E_1E_3 - 2\normofvector{p_1}\normofvector{p_3}\cos\theta_{13}
            \\
            m_{23}^2 &= m_{123}^2 + m_1^2 + m_2^2 + m_3^2 - m_{12}^2 - m_{13}^2
			\\
			&\text{$m_{12}$ is IKV directly }
        \end{aligned}
        $
        \\
        \hline
        \end{tabular}
    \caption{This set of IKVs contains four invariant mass variables and the corresponding DN is $(4;11;4322)$. 
$A$ is the phase space factor and here $\ud\Phi_4 = A \ud\alpha\ud(\cos\beta)\ud\gamma\ud m^2_{12}\ud m^2_{124} \ud m^2_{123}\ud m^2_{134}\ud (\cos\theta_{13})$ which is consistent with Eq.(\ref{eq:FORA2}).}
\end{table}


\begin{table}[H]
    \centering
	\makegapedcells
        \begin{tabular}{|c|c|}
        \hline
		$\begin{gathered}
			\textbf{IKVs}
			\\
			\textbf{DN=(3;6;3111)}
		\end{gathered}$
		&  $m_{12}^2$,$m_{13}^2$,$m_{14}^2$,$\cos\theta_{12}$,$\cos\theta_3^\star$
                \\
        \hline
        $A$ & 
    $    \frac{A_g\gammaL\betaL\abs{\boldsymbol{p_3^\star}}\abs{\boldsymbol{p_2}}}{\left(2\pi\right)^{12}2^7m\abs{\boldsymbol{p_3}}\abs{E_1\abs{\boldsymbol{p_2}}-E_2\abs{\boldsymbol{p_1}}\cos\theta_{12}}}
    $
        \\
        \hline
          $\begin{gathered}
             E_{1,\,2,\,3}
            \\
            \text{and}
            \\
             m_{12,\,13,\,23}    
        \end{gathered} $ &
        $\begin{aligned}
            E_1 &= \frac{1}{2m}\left(m_{12}^2 + m_{13}^2 + m_{14}^2 - \sum\limits_{j=1}^4 m_j^2\right)
            \\
            E_2 &= \sqrt{\normofvector{p_2}^2 + m_2^2}
            \\
            E_3 &= \gammaL\sqrt{\normofvector{p_3^\star}^2+m_3^2} - \gammaL\betaL\normofvector{p_3^\star}\cos\theta_3^\star
            \\
            m_{23}^2 &= 2m\left(E_1+E_2+E_3\right) -m^2 + \sum\limits_{j=1}^4 m_j^2 - m_{12}^2 - m_{13}^2 
\\
			&\text{$m_{12,13}$ are IKVs directly}
            \\
        \end{aligned}
        $
        \\
        \hline
        {} &
        $\begin{aligned}
            \Delta &= \lambda\left(m^2_{12},m^2_1,m^2_2\right) - 4\normofvector{p_1}^2m_2^2\sin^2\theta_{12}
            \\
            m_{134} &= \sqrt{m^2 + m_2^2 -2mE_2}
			\\
            \gammaL\betaL &= \sqrt{\gammaL^2 -1 }=\frac{\normofvector{p_2}}{m_{134}}
			\\
            \normofvector{p_3^\star} &= \frac{1}{2m_{134}}\lambda^\frac{1}{2}\left(m^2_{134},m^2_{14},m^2_3\right)
            \\
            \normofvector{p_2} &= \frac{\normofvector{p_1}\cos\theta_{12}\left(m_{12}^2-m_1^2-m_2^2\right) \pm E_1\sqrt{\Delta}}{2\left(\normofvector{p_1}^2\sin^2\theta_{12}+m_1^2\right)}
            \end{aligned}
        $
        \\
        \hline
        \end{tabular}
    \caption{This set of IKVs contains three invariant mass variables and the corresponding DN is $(3;6;3111)$. 
$A$ is the phase space factor and here $\ud\Phi_4 =A \ud\alpha\ud(\cos\beta)\ud\gamma\ud m^2_{12}\ud m^2_{13} \ud m^2_{14}\ud(\cos\theta_{12})\ud (\cos\theta_3^\star)$ which is consistent with Eq.(\ref{eq:FORA3}).
Quantities with superscripts $\star$ are defined in the rest frame of the composite particle-134.
To be precise, when $m_{12}^2 > m_1^2 + m_2^2 + 2E_1m_2$, only the positive sign in the expression of $\abs{\boldsymbol{p_2}}$ are allowed while when $m_1^2 + m_2^2 + 2\sqrt{\normofvector{p_2}^2\sin^2\theta_{12}+m_1^2}<m_{12}^2<m_1^2+m_2^2+2E_1m_2$ and $\theta_{12}<\pi/2$, both signs are allowed.
}
\label{tab:3111}
\end{table}


\begin{table}[H]
    \centering
	\makegapedcells
        \begin{tabular}{|c|c|}
        \hline
		$\begin{gathered}
			\textbf{IKVs}
			\\
			\textbf{DN=(3;6;2211)}
		\end{gathered}$
		&  $m_{12}^2$,$m_{13}^2$,$m_{24}^2$,$\cos\theta_2^\prime$,$\cos\theta_1^\star$
        \\
        \hline
       $A$ & 
    $  \frac{A_g\abs{\boldsymbol{p_1^\star}}\abs{\boldsymbol{p_2^\prime}}\lambda\left(m^2,m^2_{13},m^2_{24}\right)}{\left(2\pi\right)^{12}2^9m^3m_{13}m_{24}\abs{\boldsymbol{p_1}}\abs{\boldsymbol{p_2}}\abs{\boldsymbol{p_3}}}
    $
        \\
        \hline
          $\begin{gathered}
             E_{1,\,2,\,3}
            \\
            \text{and}
            \\
             m_{12,\,13,\,23}    
        \end{gathered} $  &
        $\begin{aligned}
            E_1 &= \gammaL\sqrt{\normofvector{p_1^\star}^2+m_1^2} - \gammaL\betaL\normofvector{p_1^\star}\cos\theta_1^\star
            \\
            E_2 &= \gammaL^\prime\sqrt{\normofvector{p_2^\prime}^2+m_2^2} - \gammaL^\prime\betaL^\prime\normofvector{p_2^\prime}\cos\theta_2^\prime
            \\
            E_3 &= \gammaL\sqrt{\normofvector{p_1^\star}^2+m_3^2} + \gammaL\betaL\normofvector{p_1^\star}\cos\theta_1^\star
            \\
            m_{23}^2 &= 2mE_2 + \sum\limits_{i=1}^4 m_i^2 - m_{12}^2 -m_{24}^2
			\\
			&\text{$m_{12,13}$ are IKVs directly}
        \end{aligned}
        $
        \\
        \hline
        {} &
        $\begin{aligned}
            \normofvector{p_1^\star} &= \frac{1}{2m_{13}}\lambda^\frac{1}{2}\left(m^2_{13},m^2_1,m^2_3\right)
            \\
            \normofvector{p_2^\prime} &= \frac{1}{2m_{24}}\lambda^\frac{1}{2}\left(m^2_{24},m^2_2,m^2_4\right)
            \\
            \gammaL\betaL &=\sqrt{\gammaL^2 -1}= \frac{1}{2mm_{13}}\lambda^\frac{1}{2}\left(m^2,m^2_{13},m^2_{24}\right)
            \\
            \gammaL^\prime\betaL^\prime &=\sqrt{{\gammaL^\prime}^2-1}= \frac{1}{2mm_{24}}\lambda^\frac{1}{2}\left(m^2,m^2_{13},m^2_{24}\right)
        \end{aligned}
        $
        \\
        \hline
        \end{tabular}
 \caption{
This set of IKVs contains three invariant mass variables and the corresponding DN is $(3;6;2211)$. 
$A$ is the phase space factor and here $\ud\Phi_4 =A \ud\alpha\ud(\cos\beta)\ud\gamma\ud m^2_{12}\ud m^2_{13} \ud m^2_{24}\ud(\cos\theta_2^\prime)\ud (\cos\theta_1^\star)$ which is consistent with Eq.(\ref{eq:FORA2}).
Quantities with superscripts $\star$ and $\prime$ are defined in the rest frame of the composite particle-13 and particle-24, respectively. 
}
\end{table}


\begin{table}[H]
    \centering
		\makegapedcells
        \begin{tabular}{|c|c|}
        \hline
		$\begin{gathered}
			\textbf{IKVs}
			\\
			\textbf{DN=(3;7;3220)}
		\end{gathered}$
		&  $m_{12}^2$,$m_{13}^2$,$m_{123}^2$,$\cos\theta_2^\star$,$\cos\theta_3^\star$
        \\
        \hline
        $A$ & 
    $   \frac{A_g\abs{\boldsymbol{p_2^\star}}\abs{\boldsymbol{p_3^\star}}\lambda\left(m^2,m^2_{123},m^2_4\right)}{\left(2\pi\right)^{12}2^9m^3m_{123}^2\abs{\boldsymbol{p_1}}\abs{\boldsymbol{p_2}}\abs{\boldsymbol{p_3}}} 
    $
        \\
        \hline
         $\begin{gathered}
             E_{1,\,2,\,3}
            \\
            \text{and}
            \\
             m_{12,\,13,\,23}    
        \end{gathered} $ &
        $\begin{aligned}
            E_1 &= \frac{1}{2m}\left(m^2 + m_{123}^2 -m_4^2\right) -2mE_2 - 2mE_3
            \\
            E_2 &= \gammaL\sqrt{\normofvector{p_2^\star}^2+m_2^2} - \gammaL\betaL\normofvector{p_2^\star}\cos\theta_2^\star
            \\
            E_3 &= \gammaL\sqrt{\normofvector{p_3^\star}^2+m_3^2} - \gammaL\betaL\normofvector{p_3^\star}\cos\theta_3^\star
            \\
            m_{23}^2 &= m_{123}^2 + m_1^2 + m_2^2 + m_3^2  - m_{12}^2 -m_{13}^2
			\\
			&\text{$m_{12,13}$ are IKVs directly}
        \end{aligned}
        $
        \\
        \hline
        {} &
        $\begin{aligned}
            \normofvector{p_2^\star} &= \frac{1}{2m_{123}}\lambda^\frac{1}{2}\left(m^2_{123},m^2_{13},m^2_2\right)
            \\
            \normofvector{p_3^\star} &= \frac{1}{2m_{123}}\lambda^\frac{1}{2}\left(m^2_{123},m^2_{12},m^2_3\right) 
            \\
            \gammaL\betaL &=\sqrt{\gammaL^2 -1 } =  \frac{1}{2mm_{123}}\lambda^\frac{1}{2}\left(m^2,m^2_{123},m^2_4\right)
        \end{aligned}
        $
        \\
        \hline
        \end{tabular}
 \caption{
This set of IKVs contains three invariant mass variables and the corresponding DN is $(3;7;3220)$. $A$ is the phase space factor and here $\ud\Phi_4 =A \ud\alpha\ud(\cos\beta)\ud\gamma\ud m^2_{12}\ud m^2_{13} \ud m^2_{123}\ud(\cos\theta_2^\star)\ud (\cos\theta_3^\star)$ which is consistent with Eq.(\ref{eq:FORA2}).
Quantities with superscripts $\star$ are defined in the rest frame of the composite particle-123.
}
\end{table}


\begin{table}[H]
    \centering
\makegapedcells
        \begin{tabular}{|c|c|}
        \hline
		$\begin{gathered}
			\textbf{IKVs}
			\\
			\textbf{DN=(3;7;3211)}
		\end{gathered}$
		&  $m_{12}^2$,$m_{24}^2$,$m_{234}^2$,$\cos\theta_2^\prime$,$\cos\theta_3^\star$
        \\
        \hline
        $A$ & 
    $   \frac{A_g\gammaL\betaL\gammaL^\prime\betaL^\prime\abs{\boldsymbol{p_3^\star}}\abs{\boldsymbol{p_2^\prime}}}{\left(2\pi\right)^{12}2^7m\abs{\boldsymbol{p_1}}\abs{\boldsymbol{p_2}}\abs{\boldsymbol{p_3}}} 
    $
        \\
        \hline
        $\begin{gathered}
             E_{1,\,2,\,3}
            \\
            \text{and}
            \\
             m_{12,\,13,\,23}    
        \end{gathered} $ &
        $\begin{aligned}
            E_1 &= \frac{1}{2m} \left(m^2 + m_1^2 - m_{234}^2\right)
            \\
            E_2 &= \gammaL^\prime\sqrt{\normofvector{p_2^\prime}^2+m_2^2} - \gammaL^\prime\betaL^\prime\normofvector{p_2^\prime}\cos\theta_2^\prime
            \\
            E_3 &= \gammaL\sqrt{\normofvector{p_3^\star}^2+m_3^2} - \gammaL\betaL\normofvector{p_3^\star}\cos\theta_3^\star
            \\
            m_{13}^2 &= 2m\left(E_1 + E_3\right) + m_{24}^2 - m^2
            \\
            m_{23}^2 &= 2mE_2 + \sum\limits_{i=1}^4 m_i^2 - m_{12}^2 - m_{24}^2
			\\
			&\text{$m_{12}$ is IKV directly}
        \end{aligned}
        $
        \\
        \hline
        {} &
        $\begin{aligned}
            \normofvector{p_3^\star} &= \frac{1}{2m_{234}}\lambda^\frac{1}{2}\left(m^2_{234},m^2_{24},m^2_3\right)
			\\
            \normofvector{p_2^\prime} &= \frac{1}{2m_{24}}\lambda^\frac{1}{2}\left(m^2_{24},m^2_{2},m^2_4\right)
            \\
            \gammaL^\prime\betaL^\prime &=\sqrt{{\gammaL^\prime}^2 - 1}= \sqrt{(\frac{m-E_1-E_3}{m_{24}})^2-1}
            \\
            \gammaL\betaL &= \sqrt{\gammaL^2 -1 } =\frac{1}{2mm_{234}}\lambda^\frac{1}{2}\left(m,m_{234},m_1\right)
            \\
        \end{aligned}
        $
        \\
        \hline
        \end{tabular}
    \caption{This set of IKVs contains three invariant mass variables and the corresponding DN is $(3;7;3221)$. 
$A$ is the phase space factor and here $\ud\Phi_4 = A \ud\alpha\ud(\cos\beta)\ud\gamma\ud m^2_{12}\ud m^2_{24} \ud m^2_{234}\ud(\cos\theta_2^\prime)\ud(\cos\theta_3^\star)$ which is consistent with Eq.(\ref{eq:FORA2}).
Quantities with superscripts $\star$ and $\prime$ are defined in the rest frame of the composite particle-234 and particle-24, respectively.
}
\end{table}


\begin{table}[H]
    \centering
\makegapedcells
        \begin{tabular}{|c|c|}
        \hline
		$\begin{gathered}
			\textbf{IKVs}
			\\
			\textbf{DN=(3;7;2221)}
		\end{gathered}$
		&  $m_{13}^2$,$m_{24}^2$,$m_{234}^2$,$\cos\theta_2^\star$,$\cos\theta_{12}$
        \\
        \hline
       $A$ & 
    $   \frac{A_g\abs{\boldsymbol{p_2^\star}}\lambda^{\frac{1}{2}}\left(m^2,m^2_{13},m^2_{24}\right)}{\left(2\pi\right)^{12}2^8m^3m_{24}\abs{\boldsymbol{p_3}}} 
    $
        \\
        \hline
       $\begin{gathered}
             E_{1,\,2,\,3}
            \\
            \text{and}
            \\
             m_{12,\,13,\,23}    
        \end{gathered} $ &
        $\begin{aligned}
            E_1 &= \frac{1}{2m} \left(m^2 + m_1^2 - m_{234}^2\right)
            \\
            E_2 &= \gammaL\sqrt{\normofvector{p_2^\star}^2+m_2^2} - \gammaL\betaL\normofvector{p_2^\star}\cos\theta_2^\star
            \\
            E_3 &= \sqrt{\frac{\lambda\left(m^2,m^2_{13},m^2_{24}\right)}{4m^2}+m_{13}^2} - E_1
            \\
            m_{12}^2 &= m_1^2 + m_2^2 + 2E_1E_2 - 2\normofvector{p_1}\normofvector{p_2}\cos\theta_{12}
            \\
            m_{23}^2 &= m^2 + \sum\limits_{i=1}^4 m_i^2 -m_{12}^2 - m_{13}^2 - 2mE_4
			\\
			&\text{$m_{12}$ is IKV directly}
        \end{aligned}
        $
        \\
        \hline
       {} &
        $\begin{aligned}
            \gammaL\betaL &=\sqrt{\gammaL^2-1}= \frac{1}{2mm_{24}}\lambda^\frac{1}{2}\left(m^2,m^2_{13},m^2_{24}\right)
			 \\
            \normofvector{p_2^\star} &= \frac{1}{2m_{24}}\lambda^\frac{1}{2}\left(m^2_{24},m^2_2,m^2_4\right)
            \\
            E_4 &= \gammaL\sqrt{\normofvector{p_2^\star}^2+m_4^2} + \gammaL\betaL\normofvector{p_2^\star}\cos\theta_2^\star
            \\
        \end{aligned}
        $
        \\
        \hline
        \end{tabular}
 \caption{
 This set of IKVs contains three invariant mass variables and the corresponding DN is $(3;7;2221)$. 
$A$ is the phase space factor and here $\ud\Phi_4 = A \ud\alpha\ud(\cos\beta)\ud\gamma\ud m^2_{13}\ud m^2_{24} \ud m^2_{234}\ud(\cos\theta_2^\star)\ud(\cos\theta_{12})$ which is consistent with Eq.(\ref{eq:FORA2}).
Quantities with superscripts $\star$ are defined in the rest frame of the composite particle-24.}
\end{table}


 \begin{table}[H]
    \centering
\makegapedcells
        \begin{tabular}{|c|c|}
        \hline
		$\begin{gathered}
			\textbf{IKVs}
			\\
			\textbf{DN=(3;8;3221)}
		\end{gathered}$
		&  $m_{13}^2$,$m_{134}^2$,$m_{234}^2$,$\cos\theta_4^\star$,$\cos\theta_{12}$
        \\
        \hline
        $A$ & 
    $   \frac{A_g\abs{\boldsymbol{p_4^\star}}\lambda^{\frac{1}{2}}\left(m^2,m^2_{2},m^2_{134}\right)}{\left(2\pi\right)^{12}2^8m^3m_{134}\abs{\boldsymbol{p_3}}}
    $
        \\
        \hline
        $\begin{gathered}
             E_{1,\,2,\,3}
            \\
            \text{and}
            \\
             m_{12,\,13,\,23}    
        \end{gathered} $ &
         $\begin{aligned}
            E_1 &= \frac{1}{2m} \left(m^2 + m_1^2 - m_{234}^2\right)
            \\
            E_2 &= \frac{1}{2m} \left(m^2 + m_2^2 - m_{134}^2\right)
            \\
            E_3 &= m- E_1 - E_2 - E_3
            \\
            m_{12}^2 &= m_1^2 + m_2^2 + 2E_1E_2 - 2\normofvector{p_1}\normofvector{p_2}\cos\theta_{12}
            \\
            m_{23}^2 &= m^2 + \sum\limits_{i=1}^4 m_i^2 -m_{12}^2 - m_{13}^2 - 2mE_4
			\\
			&\text{$m_{13}$ is IKV directly}
        \end{aligned}
        $
        \\
        \hline
        {} &
        $\begin{aligned}
            \gammaL\betaL &= \sqrt{\gammaL^2 -1 } = \frac{1}{2mm_{134}}\lambda^\frac{1}{2}\left(m^2,m^2_{134},m^2_{2}\right)
            \\
            \normofvector{p_4^\star} &= \frac{1}{2m_{134}}\lambda^\frac{1}{2}\left(m^2_{134},m^2_{13},m^2_4\right)
            \\
            E_4 &= \gammaL\sqrt{\normofvector{p_4^\star}^2+m_4^2} - \gammaL\betaL\normofvector{p_4^\star}\cos\theta_4^\star
            \\
        \end{aligned}
        $
        \\
        \hline
        \end{tabular}
    \caption{This set of IKVs contains three invariant mass variables and the corresponding DN is $(3;8;3221)$. 
$A$ is the phase space factor and here $\ud\Phi_4 = A\ud\alpha\ud(\cos\beta)\ud\gamma\ud m^2_{13}\ud m^2_{134} \ud m^2_{234}\ud(\cos\theta_4^\star)\ud(\cos\theta_{12})$ which is consistent with Eq.(\ref{eq:FORA2})
Quantities with superscripts $\star$ are defined in the rest frame of the composite particle-134.}
\end{table}


\begin{table}[H]
    \centering
\makegapedcells
        \begin{tabular}{|c|c|}
        \hline
		$\begin{gathered}
			\textbf{IKVs}
			\\
			\textbf{DN=(3;8;3311)}
		\end{gathered}$
		&  $m_{12}^2$,$m_{123}^2$,$m_{124}^2$,$\cos\theta_1^\star$,$\cos\theta_{13}$
        \\
        \hline
       $A$ & 
    $  \frac{A_g\gammaL\betaL\abs{\boldsymbol{p_1^\star}}}{\left(2\pi\right)^{12}2^7m^2\abs{\boldsymbol{p_2}}}
    $
        \\
        \hline
        $\begin{gathered}
             E_{1,\,2,\,3}
            \\
            \text{and}
            \\
             m_{12,\,13,\,23}    
        \end{gathered} $  &
        $\begin{aligned}
            E_1 &= \gammaL\sqrt{\normofvector{p_1^\star}^2+m_1^2} - \gammaL\betaL\normofvector{p_1^\star}\cos\theta_1^\star
            \\
            E_2 &= \gammaL\sqrt{\normofvector{p_1^\star}^2+m_2^2} + \gammaL\betaL\normofvector{p_1^\star}\cos\theta_1^\star
            \\
            E_3 &= \frac{1}{2m} \left(m^2 + m_3^2 - m_{124}^2\right)
            \\
            m_{13}^2 &= m_1^2 + m_3^2 + 2E_1E_3 - 2\normofvector{p_1}\normofvector{p_3}\cos\theta_{13}
            \\
            m_{23}^2 &= m^2 + \sum\limits_{i=1}^4 m_i^2 -m_{12}^2 - m_{13}^2 - 2mE_4
			\\
			&\text{$m_{12}$ is IKV directly}
        \end{aligned}
        $
        \\
        \hline
        {} &
        $\begin{aligned}
            \gammaL\betaL &=\sqrt{\gammaL^2 -1 }= \sqrt{\frac{\left(m_{123}^2 + m_{124}^2 - m_3^2 - m_4^2\right)^2}{4m^2m^2_{12}}-1}
            \\
            E_4 &= \frac{1}{2m}\left(m^2 + m_4^2 - m_{123}^2\right)
			\\
            \normofvector{p_1^\star} &= \frac{1}{2m_{12}}\lambda^\frac{1}{2}\left(m^2_{12},m^2_{1},m^2_2\right)
            \\
        \end{aligned}
        $
        \\
        \hline
        \end{tabular}
    \caption{This set of IKVs contains three invariant mass variables and the corresponding DN is $(3;8;3311)$. 
$A$ is the phase space factor and here $\ud\Phi_4 = A\ud\alpha\ud(\cos\beta)\ud\gamma\ud m^2_{12}\ud m^2_{123} \ud m^2_{124}\ud(\cos\theta_1^\star)\ud(\cos\theta_{13})$ which is consistent with Eq.(\ref{eq:FORA2}).
Quantities with superscripts $\star$ are defined in the rest frame of the composite particle-12.}
\end{table}


\begin{table}[H]
    \centering
\makegapedcells
        \begin{tabular}{|c|c|}
        \hline
		$\begin{gathered}
			\textbf{IKVs}
			\\
			\textbf{DN=(3;9;3222)}
		\end{gathered}$
		&  $m_{124}^2$,$m_{134}^2$,$m_{234}^2$,$\cos\theta_{12}$,$\cos\theta_{13}$
        \\
        \hline
        $A$ & 
    $  \frac{A_g\abs{\boldsymbol{p_1}}}{\left(2\pi\right)^{12}2^7m^3}
    $
        \\
        \hline
         $\begin{gathered}
             E_{1,\,2,\,3}
            \\
            \text{and}
            \\
             m_{12,\,13,\,23}    
        \end{gathered} $  &
        $\begin{aligned}
           E_1 &= \frac{1}{2m}\left(m^2 + m_1^2 - m_{234}^2\right)
            \\
           E_2 &= \frac{1}{2m}\left(m^2 + m_2^2 - m_{134}^2\right)
            \\
           E_3 &= \frac{1}{2m}\left(m^2 + m_3^2 - m_{124}^2\right)
            \\
            m_{12}^2 &= m_1^2 + m_2^2 + 2E_1E_2 - 2\normofvector{p_1}\normofvector{p_2}\cos\theta_{12} 
            \\
            m_{13}^2 &= m_1^2 + m_3^2 + 2E_1E_3 - 2\normofvector{p_1}\normofvector{p_3}\cos\theta_{13}
            \\
            m_{23}^2 &= -m^2 + \sum\limits_{i=1}^4 m_i^2 -m_{12}^2 - m_{13}^2 + 2m \left( E_1+E_2+E_3 \right)
        \end{aligned}
        $
        \\
        \hline
        \end{tabular}
    \caption{This set of IKVs contains three invariant mass variables and the corresponding DN is $(3;9;3222)$. 
$A$ is the phase space factor and here $\ud\Phi_4 = A\ud\alpha\ud(\cos\beta)\ud\gamma\ud m^2_{124}\ud m^2_{134} \ud m^2_{234}\ud(\cos\theta_{12})\ud(\cos\theta_{13})$ which is consistent with Eq.(\ref{eq:FORA}).
$\theta_{12}$ and $\theta_{13}$ denote the angles between three-momenta of the particle 1 and particle 2 and that of the particle 1 and particle 3, repsectively.}

\end{table}

\begin{table}[H]
    \centering
\makegapedcells
        \begin{tabular}{|c|c|}
        \hline
		$\begin{gathered}
			\textbf{IKVs}
			\\
			\textbf{DN=(2;4;1111)}
		\end{gathered}$
		&  $m_{13}^2$,$m_{24}^2$,$\cos\theta_1^\star$,$\cos\theta_{12}$,$\cos\theta_{13}$
        \\
        \hline
        $A$ & 
    $  \frac{A_g\gammaL\betaL\abs{\boldsymbol{p_1^\star}}}{\left(2\pi\right)^{12}2^7m^2\abs{\boldsymbol{p_2}}}
    $
        \\
        \hline
         $\begin{gathered}
             E_{1,\,2,\,3}
            \\
            \text{and}
            \\
             m_{12,\,13,\,23}    
        \end{gathered} $ &
        $\begin{aligned}
            E_1 &= \gammaL^\star\sqrt{\normofvector{p_1^\star}^2+m_1^2}-\gammaL^\star\betaL^\star\normofvector{p_1^\star}\cos\theta_1^\star
            \\
            E_2 &= \gammaL^\prime\sqrt{\normofvector{p_2^\prime}^2+m_2^2} - \gammaL^\prime\betaL^\prime\normofvector{p_2^\prime}\cos\theta_2^\prime
            \\
            E_3 &= \gammaL^\star\sqrt{\normofvector{p_1^\star}^2+m_3^2}+\gammaL^\star\betaL^\star\normofvector{p_1^\star}\cos\theta_1^\star
            \\
            m_{12}^2 &= m_1^2 + m_2^2 + 2E_1E_2 - 2\normofvector{p_1}\normofvector{p_2}\cos\theta_{12} 
            \\
            m_{23}^2 &= m^2 + \sum\limits_{i=1}^4 m_i^2 -m_{12}^2 - m_{13}^2 - 2mE_4
        \end{aligned}
        $
        \\
        \hline
        {} &
        $\begin{aligned}
            \gammaL\betaL &=\sqrt{\gammaL^2 -1 }= \frac{1}{2mm_{13}}\lambda^\frac{1}{2}\left(m^2,m^2_{13},m^2_{24}\right)
            \\
            \gammaL^\prime\betaL^\prime &=\sqrt{{\gammaL^\prime}^2 -1 }= \frac{1}{2mm_{24}}\lambda^\frac{1}{2}\left(m^2,m^2_{13},m^2_{24}\right)
            \\
            \normofvector{p_1^\star} &= \frac{1}{2m_{13}}\lambda^\frac{1}{2}\left(m^2_{13},m^2_1,m^2_3\right)
			\\
            \normofvector{p_2^\prime} &=\frac{1}{2m_{24}}\lambda^\frac{1}{2}\left(m^2_{24},m^2_2,m^2_4\right)
            \\
            E_4 &= \gammaL^\prime\sqrt{\normofvector{p_2^\prime}^2+m_4^2} + \gammaL^\prime\betaL^\prime\normofvector{p_2^\prime}\cos\theta_2^\prime
        \end{aligned}
        $
        \\
        \hline
        \end{tabular}
\caption{
This set of IKVs contains two invariant mass variables and the corresponding DN is $(2;4;1111)$. 
$A$ is the phase space factor and here $\ud\Phi_4 = A \ud\alpha\ud(\cos\beta)\ud\gamma\ud m^2_{13}\ud m^2_{24} \ud(\cos\theta_1^\star) \ud(\cos\theta_{12}\ud \cos\theta_{13})$ which is consistent with Eq.(\ref{eq:FORA2}).
Quantities with superscripts $\star$ and $\prime$ are defined in the rest frame of the composite particle-13 and particle-24, respectively. }
 \end{table}

 \begin{table}[H]
        \centering
\makegapedcells
        \begin{tabular}{|c|c|}
        \hline
		$\begin{gathered}
			\textbf{IKVs}
			\\
			\textbf{DN=(2;4;2110)}
		\end{gathered}$
		&  $m_{12}^2$,$m_{13}^2$,$\cos\theta_1^\star$,$\theta_{34}$,$\theta_{4(12)}$
            \\
            \hline
           $A$ & 
        $  \frac{A_g\abs{\boldsymbol{p_1^\star}}\abs{\boldsymbol{p_4}}\abs{\boldsymbol{p_1}+\boldsymbol{p_2}}}{\left(2\pi\right)^{12}2^6m_{12}\abs{\boldsymbol{p_1}}\abs{\boldsymbol{p_2}}} \frac{\normofvector{p_3}\normofvector{p_4}\sin^2\theta_{34}}{\left(E_4E_{12}\sin^2\theta_{4(12)} + E_3E_{12}\sin^2(\theta_{34}+\theta_{4(12)}) + E_3E_4\sin^2\theta_{34}\right)}
        $
            \\
            \hline
            $\begin{gathered}
             E_{1,\,2,\,3}
            \\
            \text{and}
            \\
             m_{12,\,13,\,23}    
        \end{gathered} $  &
            $\begin{aligned}
               E_1 &= \gammaL\sqrt{\normofvector{p_1^\star}^2+m_1^2}-\gammaL\betaL\normofvector{p_1^\star}\cos\theta_1^\star
                \\
               E_2 &= \gammaL\sqrt{\normofvector{p_1^\star}^2+m_1^2}+\gammaL\betaL\normofvector{p_1^\star}\cos\theta_1^\star
                \\
               E_3 &= \sqrt{\frac{\sin^2\theta_{4(12)}}{\sin^2\theta_{34}}\normofvector{p_1+p_2}^2 + m_3^2}
                \\
                m_{23}^2 &= -m^2 + \sum\limits_{i=1}^4 m_i^2 -m_{12}^2 - m_{13}^2 + 2m\left(E_1+E_2+E_3\right)
				\\
				&\text{$m_{12,13}$ are IKVs directly }
            \end{aligned}
            $
            \\
            \hline
            {} &
            $\begin{aligned}
                \gammaL\betaL &=\sqrt{\gammaL^2 -1}= \frac{\normofvector{p_1+p_2}}{m_{12}}
                \\
                \normofvector{p_1^\star} &= \frac{1}{2m_{12}} \lambda^\frac{1}{2} \left(m^2_{12},m^2_1,m^2_2\right)
                \\
             &\normofvector{p_1+p_2} \text{is solved from equation,}
             \\
                 \sqrt{\frac{\sin^2\theta_{4(12)}}{\sin^2\theta_{34}}\normofvector{p_1+p_2}^2 + m_3^2}  &+ \sqrt{\frac{\sin\left(\theta_{34}+\theta_{4(12)}\right)^2}{\sin^2\theta_{34}}\normofvector{p_1+p_2}^2+m_4^2} + \sqrt{\normofvector{p_1+p_2}^2 + m_{12}^2} = m
            \end{aligned}
            $
            \\
            \hline
            \end{tabular}
\caption{
This set of IKVs contains two invariant mass variables and the corresponding DN is $(2;4;2110)$. 
$A$ is the phase space factor and here $\ud\Phi_4= A\ud\alpha\ud(\cos\beta)\ud\gamma\ud m^2_{12}\ud m^2_{13} \ud(\cos\theta_1^\star) \ud\theta_{34}\ud\theta_{4(12)}$ which is consistent with Eq.(\ref{eq:FORA2}).
Quantities with superscripts $\star$ are defined in the rest frame of the composite particle-12.
}        
\end{table}

\begin{table}[H]
    \centering
\makegapedcells
        \begin{tabular}{|c|c|}
        \hline
		$\begin{gathered}
			\textbf{IKVs}
			\\
			\textbf{DN=(2;5;2111)}
		\end{gathered}$
		&  $m_{12}^2$,$m_{234}^2$,$\cos\theta_3^\star$,$\cos\theta_{12}$,$\cos\theta_{13}$
        \\
        \hline
       $A$ & 
    $  \frac{A_g\gammaL\betaL\abs{\boldsymbol{p_3^\star}}\abs{\boldsymbol{p_2}}\abs{\boldsymbol{p_1}}}{\left(2\pi\right)^{12}2^6m\abs{E_1\abs{\boldsymbol{p_2}}-E_2\abs{\boldsymbol{p_1}}\cos\theta_{12}}}
    $
        \\
        \hline
        $\begin{gathered}
             E_{1,\,2,\,3}
            \\
            \text{and}
            \\
             m_{12,\,13,\,23}    
        \end{gathered} $  &
        $\begin{aligned}
           E_1 &= \frac{1}{2m}\left(m^2 + m_1^2 - m_{234}^2\right)
            \\
           E_2 &= \sqrt{\left(\frac{\normofvector{p_1}\cos\theta_{12}\left(m_{12}^2-m_1^2-m_2^2\right) \pm E_1^0\sqrt{\Delta}}{2\left(\normofvector{p_1}^2\sin^2\theta_{12}+m_1^2\right)}\right)^2 + m_2^2}
            \\
           E_3 &= \gammaL\sqrt{\normofvector{p_3^\star}^2+m_3^2} - \gammaL\betaL\normofvector{p_3^\star}\cos\theta_3^\star
            \\
            m_{13}^2 &= m_1^2 + m_3^2 + 2E_1E_3 - 2\normofvector{p_1}\normofvector{p_3}\cos\theta_{13}
            \\
            m_{23}^2 &= -m^2 + \sum\limits_{i=1}^4 m_i^2 -m_{12}^2 - m_{13}^2 + 2m\left(E_1+E_2+E_3\right) 
			\\
			&\text{$m_{12}$ is IKV directly}        
        \end{aligned}
        $
        \\
        \hline
        {} &
        $\begin{aligned}
            \Delta &= \lambda\left(m^2_{12},m^2_1,m^2_2\right) - 4\normofvector{p_1}^2m_2^2\sin^2\theta_{12}
            \\
            m_{34}^2 &= m^2 + m_{12}^2 - 2m\left(E_1+E_2\right)
            \\
            \gammaL\betaL &=\sqrt{\gammaL^2 -1 }= \sqrt{(\frac{m-E_1-E_2}{m_{34}})^2 - 1}
            \\
            \normofvector{p_3^\star} &= \frac{1}{2m_{34}}\lambda^\frac{1}{2}\left(m^2_{34},m^2_{4},m^2_3\right)
            \\
        \end{aligned}
        $
        \\
        \hline
        \end{tabular}
    \caption{This set of IKVs contains two invariant mass variables and the corresponding DN is $(2;5;2111)$. 
$A$ is the phase space factor and here $\ud\Phi_4 = A\ud\alpha\ud(\cos\beta)\ud\gamma\ud m^2_{12}\ud m^2_{234} \ud(\cos\theta_3^\star) \ud\theta_{12}\ud\theta_{13}$ which is consistent with Eq.(\ref{eq:FORA3}).
Quantities with superscripts $\star$ are defined in the rest frame of the composite particle-34.
To be precise, when $m_{12}^2 > m_1^2 + m_2^2 + 2E_1m_2$, only the positive sign in the expression of $\abs{\boldsymbol{p_2}}$ are allowed while when $m_1^2 + m_2^2 + 2\sqrt{\normofvector{p_2}^2\sin^2\theta_{12}+m_1^2}<m_{12}^2<m_1^2+m_2^2+2E_1m_2$ and $\theta_{12}<\pi/2$, both signs are allowed.
}
\label{tab:2111}    
\end{table}


\begin{table}[H]
    \centering
\makegapedcells
        \begin{tabular}{|c|c|}
        \hline
		$\begin{gathered}
			\textbf{IKVs}
			\\
			\textbf{DN=(2;6;2211)}
		\end{gathered}$
		&  $m_{134}^2$,$m_{234}^2$,$\cos\theta_3^\star$,$\cos\theta_{12}$,$\cos\theta_{13}$
        \\
        \hline
        $A$ & 
    $   \frac{A_g\gammaL\betaL\abs{\boldsymbol{p_3^\star}}\abs{\boldsymbol{p_1}}}{\left(2\pi\right)^{12}2^6m^2}
    $
        \\
        \hline
        $\begin{gathered}
             E_{1,\,2,\,3}
            \\
            \text{and}
            \\
             m_{12,\,13,\,23}    
        \end{gathered} $  &
        $\begin{aligned}
            E_1 &= \frac{1}{2m}\left(m^2+m_1^2-m_{234}^2\right) 
            \\
            E_2 &= \frac{1}{2m}\left(m^2+m_2^2-m_{134}^2\right)
            \\
            E_3 &= \gammaL\sqrt{\normofvector{p_3^\star}^2+m_3^2}-\gammaL\betaL\normofvector{p_3^\star}\cos\theta_3^\star
            \\
            m_{12}^2 &= m_1^2 + m_2^2 + 2E_1E_2 - 2\normofvector{p_1}\normofvector{p_2}\cos\theta_{1,2} 
            \\
            m_{13}^2 &= m_1^2+m_3^2+2E_1E_3-2\normofvector{p_1}\normofvector{p_3}\cos\theta_{1,3}
            \\
            m_{23}^2 &= m^2 + \sum\limits_{i=1}^4 m_i^2 -m_{12}^2 - m_{13}^2 -2mE_4
        \end{aligned}
        $
        \\
        \hline
       {} &
        $\begin{aligned}
            m_{34}^2 &= m_{134}^2 + m_{234}^2 + m_{12}^2 - m_1^2 - m_2^2 - m^2
            \\
            \gammaL\betaL &= \sqrt{\gammaL^2 - 1} = \sqrt{\frac{ \left(m_{234}^2 + m_{134}^2 -m_1^2 - m_2^2\right)^2}{4m^2m^2_{34}}-1}
            \\
            \normofvector{p_3^\star} &= \frac{1}{2m_{34}}\lambda^\frac{1}{2}\left(m^2_{34},m^2_3,m^2_4\right)
            \\
            E_4 &= \gamma_{L}\sqrt{\normofvector{p_3^\star}^2+m_4^2}+\gammaL\betaL\normofvector{p_3^\star}\cos\theta_3^\star
            \\
        \end{aligned}
        $
        \\
        \hline
        \end{tabular}
    \caption{This set of IKVs contains two invariant mass variables and the corresponding DN is $(2;6;2211)$. 
$A$ is the phase space factor and here $\ud\Phi_4 = A \ud\alpha\ud(\cos\beta)\ud\gamma\ud m^2_{134}\ud m^2_{234} \ud(\cos\theta_3^\star) \ud\theta_{12}\ud\theta_{13}$ which is consistent with Eq.(\ref{eq:FORA2}).
Quantities with superscripts $\star$ are defined in the rest frame of the composite particle-34.
}
    \end{table}


\begin{table}[H]
    \centering
\makegapedcells
        \begin{tabular}{|c|c|}
        \hline
		$\begin{gathered}
			\textbf{IKVs}
			\\
			\textbf{DN=(1;2;1100)}
		\end{gathered}$
		&  $m_{12}^2$,$\cos\theta_{1}^\star$,$\cos\theta_{13}$,$\theta_{34}$,$\theta_{4(12)}$
        \\
        \hline
        $A$ & 
    $  \frac{A_g\abs{\boldsymbol{p_1}+\boldsymbol{p_2}}\normofvector{p_1^\star}\normofvector{p_3}^2\normofvector{p_4}^2}{\left(2\pi\right)^{12}2^5m_{12}\abs{\boldsymbol{p_2}}} \frac{\sin^2\theta_{34}}{\left(E_4E_{12}\sin^2\theta_{4(12)} + E_3E_{12}\sin^2(\theta_{34}+\theta_{4(12)})+E_3E_4\sin^2\theta_{34}\right)}
    $
        \\
        \hline
       $\begin{gathered}
             E_{1,\,2,\,3}
            \\
            \text{and}
            \\
             m_{12,\,13,\,23}    
        \end{gathered} $ &
        $\begin{aligned}
           E_1 &= \gammaL\sqrt{\normofvector{p_1^\star}^2+m_1^2}-\gammaL\betaL\normofvector{p_1^\star}\cos\theta_1^\star
            \\
           E_2 &= \gammaL\sqrt{\normofvector{p_1^\star}^2+m_2^2}+\gammaL\betaL\normofvector{p_1^\star}\cos\theta_1^\star 
            \\
           E_3 &= \sqrt{\frac{\sin^2\theta_{4(12)}}{\sin^2\theta_{34}}\normofvector{p_{1}+p_2}^2 + m_3^2}
            \\
            m_{13}^2 &= m_1^2+m_3^2+2E_1E_3-2\normofvector{p_1}\normofvector{p_3}\cos\theta_{13}
            \\
            m_{23}^2 &= m^2 + \sum\limits_{i=1}^4 m_i^2 -m_{12}^2 - m_{13}^2 -2mE_4
            \\
			&\text{$m_{12}$ is IKV directly}
    \end{aligned}
        $
        \\
        \hline
        {} &
        $\begin{aligned}
            \gammaL\betaL &=\sqrt{\gammaL^2 -1 }= \frac{\normofvector{p_1 + p_2}}{m_{12}}
            \\
            \normofvector{p_1^\star} &= \frac{1}{2m_{12}} \lambda^\frac{1}{2} \left(m^2_{12},m^2_1,m^2_2\right)
            \\
           &\normofvector{p_{1}+p_2} \text{is solved from equation}
            \\
           \sqrt{\frac{\sin^2\theta_{4(12)}}{\sin^2\theta_{34}}\normofvector{p_{1}+p_2}^2  + m_3^2} &+ \sqrt{\frac{\sin^2\left(\theta_{34}+\theta_{4(12)}\right)}{\sin^2\theta_{34}}\normofvector{p_1+p_2}^2+m_4^2} + \sqrt{\normofvector{p_1+p_2}^2 + m_{12}^2} = m
        \end{aligned}
        $
        \\
        \hline
        \end{tabular}
 \caption{
 This set of IKVs contains one invariant mass variables and the corresponding DN is $(1;2;1100)$. 
$A$ is the phase space factor and here $\ud\Phi_4 = A \ud\alpha\ud(\cos\beta)\ud\gamma\ud m^2_{12}\ud(\cos\theta_1^\star)\ud(\cos\theta_{13})\ud\theta_{34}\ud\theta_{4(12)}$ which is consistent with Eq.(\ref{eq:FORA2}).
Quantities with superscripts $\star$ are defined in the rest frame of the composite particle-12.
}
    \end{table}

    \begin{table}[H]
        \centering
\makegapedcells
        \begin{tabular}{|c|c|}
        \hline
		$\begin{gathered}
			\textbf{IKVs}
			\\
			\textbf{DN=(1;3;1110)}
		\end{gathered}$
		&  $m_{234}^2$,$\cos\theta_{3}^\star$,$\cos\theta_{12}$,$\theta_{13}$,$\theta_{2(34)}$
            \\
            \hline
            $A$ & 
        $ \frac{A_g\abs{\boldsymbol{p_3}+\boldsymbol{p_4}}\abs{\boldsymbol{p_3^\star}}\abs{\boldsymbol{p_1}}^2\abs{\boldsymbol{p_2}}\sin^2\theta_{12}}{\left(2\pi\right)^{12}2^5mm_{34}E_2\sin^2\theta_{2(34)}}
        $
            \\
            \hline
           $\begin{gathered}
             E_{1,\,2,\,3}
            \\
            \text{and}
            \\
             m_{12,\,13,\,23}    
        \end{gathered} $ &
            $\begin{aligned}
                E_1 &= \frac{1}{2m} \left(m^2 + m_1^2 - m_{234}^2\right)
                \\
                E_2 &= \sqrt{\normofvector{p_2}^2 + m_2^2}
                \\
                E_3 &= \gammaL\sqrt{\normofvector{p_3^\star}^2+m_3^2}-\gammaL\betaL\normofvector{p_3^\star}\cos\theta_3^\star
                \\
                m_{12}^2 &= m_1^2 + m_2^2 + 2E_1E_2 - 2\normofvector{p_1}\normofvector{p_2}\cos\theta_{12}
                \\
                m_{13}^2 &= m_1^2+m_3^2+2E_1E_3-2\normofvector{p_1}\normofvector{p_3}\cos\theta_{13}
                \\
                m_{23}^2 &= m^2 + \sum\limits_{i=1}^4 m_i^2 -m_{12}^2 - m_{13}^2 -2mE_4
                \\
            \end{aligned}
            $
            \\
            \hline
           {} &
            $\begin{aligned}
                \gammaL\betaL &=\sqrt{\gammaL^2 -1 }=\frac{\normofvector{p_3+p_4}}{m_{34}}
                \\
                m_{34}^2 &= m^2 + m_{12}^2 - 2m\left(E_1+E_2\right)
                \\
                \normofvector{p_3^\star} &= \frac{1}{2m_{34}}\lambda^\frac{1}{2}\left(m^2_{34},m^2_{3},m^2_4\right)
                \\
            &\normofvector{p_3+p_4} \text{is solved from equation,}
            \\
                 \sqrt{\frac{\sin^2\theta_{2(34)}}{\sin^2\theta_{12}}\normofvector{p_3+p_4}^2 + m_{1}^2} &
                 + \sqrt{\frac{\sin^2\left(\theta_{2(34)}+\theta_{12}\right)}{\sin^2\theta_{2(34)}}\frac{\sin^2\theta_{2(34)}}{\sin^2\theta_{12}}\normofvector{p_3+p_4}^2+m_2^2} 
                 + \sqrt{\normofvector{p_3+p_4}^2 + m_{34}^2} = m
            \end{aligned}
            $
            \\
            \hline
            \end{tabular}
\caption{
This set of IKVs contains one invariant mass variables and the corresponding DN is $(1;3;1110)$. 
$A$ is the phase space factor and here $\ud\Phi_4  = A\ud\alpha\ud(\cos\beta)\ud\gamma\ud m^2_{234}\ud(\cos\theta_1^\star)\ud(\cos\theta_{12})\ud\theta_{13}\ud\theta_{2(34)}$ which is consistent with Eq.(\ref{eq:FORA2}). 
Quantities with superscripts $\star$ are defined in the rest frame of the composite particle-34. 
}
\label{tab:1110}
\end{table}

\section{Application for the reaction $e+p\to e+J/\psi+p$}
\label{sec5}


Now Monte-Carlo method has been widely used in the numerical calculation of $n$-body final states process.
However, once the extremely sharp peak structure appears in the amplitude, the efficiency of the Monte-Carlo method will decrease since significances of the sample points are required to guarantee the precision.  
Nevertheless, the explicit formulae listed here will avoid this problem.
%
For example, if the photon is an intermediate state and the invariant mass can be very close to zero, there will be a sharp structure because of the photon's propagator.
At that time, the Monte-Carlo method needs to be improved, such as the adaptive Monte-Carlo method.
But if we use the exact equations shown here, the usual numerical method is enough to finish the calculation.
Here we give an example to show how to distinguish the signal of $P_c$ states and the background of Permeron exchange in the reaction of $e+p\to e+J/\psi+p$.
%
With formulae given in this paper and Eqs.(\ref{eq:dsigma}) and (\ref{eq:dphi}), one can calculate the $\ud\sigma$ straightforwardly.

\subsection{Background}

%
There are three $P_c$ states identified from analyzing the $J/\psi-p$ invariant mass distributions of the decay process $\Lambda^*_c\to K J/\psi p$ measured by the LHCb Collaboration in 2019 \cite{LHCb:2019kea}.
However, these pentaquark resonance signals are only observed at LHCb so far. 
Thus, it is of great importance to reconfirm the pentaquark resonance at other experiments. 
These $P_c$ states can also be investigated by using the electromagnetic production of $J/\psi$ from the nucleon, such as $e+p \to e+J/\psi + p$ studied in Ref.~\cite{Wu:2010jy}.
As shown in Ref.\cite{GlueX:2019mkq}, the GlueX Collaboration did not find the evidence for $P_c$ states, although the statistic is not very high and it just collected around 500 events for $J/\psi$ in the all phase space.
As discussed in Ref.\cite{Wu:2019adv}, because of the large background of Pomeron exchange mechanism, the pure signal of $P_c$ states can only be clear around the forward angle of outgoing $J/\psi$.
On the other hand, in Ref.\cite{Yang:2020eye}, the ratio of the signal to background would increase significantly with proper kinematic cut for the $e+p \to e+J/\psi + p$ reaction.
Furthermore, the EicC has a higher signal over background ratio than that of the JLab12.
Here, as an example of application of the formulae given in this paper, an analysis of the $e+p \to e+J/\psi + p$ reaction at the energy of the EicC experiment is performed below.
Because it is a collision process with three-body final state, there are four IKVs at least. 
After integrating one variable, a three-dimension distribution plot will be shown to distinguish the background and pentaquark states.

\subsection{Mechanism}

\begin{figure}[H]
\centering
\begin{subfigure}[h]{0.48\textwidth}
\centering
\includegraphics[scale=0.7]{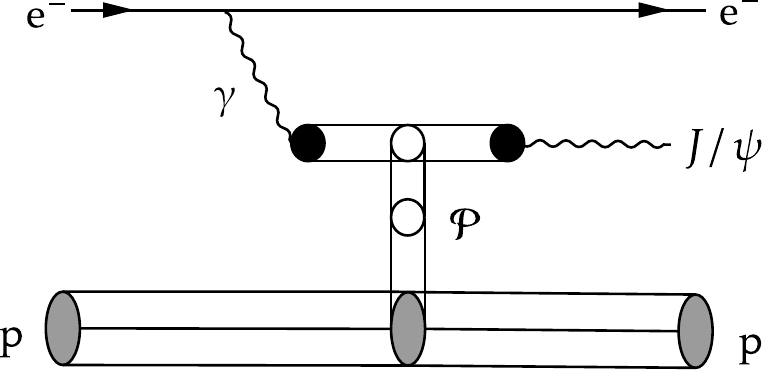}
\caption{}
\end{subfigure}
\hfill
\begin{subfigure}[h]{0.48\textwidth}
\centering
\includegraphics[scale=0.7]{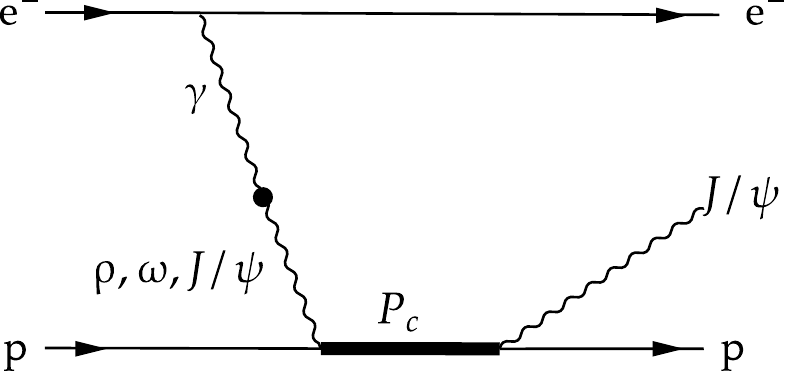}
\caption{}
\end{subfigure}
\caption{Mechanism of the interaction $e+p\rightarrow e +J/\psi+p$. (a) and (b) are diagrams of background channel and signal channel, respectively.}
\label{fig:Feynman}
\end{figure}

There are two main mechanisms for the process $e + p \to e + J/\psi + p$, namely Pomeron-exchange and $P_c$ resonance which are shown in Fig.\ref{fig:Feynman}(a) and (b), respectively. 

The amplitude $\mathcal{M}$ of the process $e+p\to e+J/\psi +p$ can be written as,
\begin{equation}\label{M}
    \mathcal{M}_{ep\rightarrow eVp} = \mathcal{M}^\mu_{R_1}\frac{-g_{\mu\nu}}{q^2}\mathcal{M}^\nu_{R_2},
\end{equation}
where $\mathcal{M}^\mu_{R_1}$ and $\mathcal{M}^\nu_{R_2}$ denote the amplitudes of subprocess $e\rightarrow e\gamma$ and $\gamma p\rightarrow pJ/\psi$, respectively.
The $\mathcal{M}^\mu_{R_1}$ can be obtained straightforwardly from the quantum electrodynamic theory, 
\begin{align}
    \mathcal{M}^\mu_{R_1}=ie\bar{u}_e(k^\prime,\lambda^\prime)\gamma^\mu u_e(k,\lambda),\label{MR1}
\end{align}
where $u_e$ is the spinor of electron, 
$k(k^\prime)$ and $\lambda(\lambda^\prime)$ are the four-momentum and the z-component of the spin for incoming (outgoing) electron, respectively.
On the other hand, $\mathcal{M}^\nu_{R_2}$ includes two parts corresponding to the different mechanisms,
\begin{align}
    \mathcal{M}_{R_2}^{\nu}&=
    \bar{u}_{p}\left(p^{\prime}, m_{s}^{\prime}\right) \epsilon_{\mu}^{*}\left(q^{\prime}, \lambda_{J / \psi}^{\prime}\right)
\left( 
\mathcal{M}^{\mu \nu}_{\mathbb{P}}\left(q, p, q^{\prime}, p^{\prime}\right) 
+
\mathcal{M}^{\mu \nu}_{P_c}\left(q, p, q^{\prime}, p^{\prime}\right) 
\right)
u_{p}\left(p, m_{s}\right),\label{MR2}
\end{align}
where $q$ is the four-momentum of intermediate photon and $p(m_s)$, $q^\prime(\lambda^\prime_{J/\psi})$, and $p^\prime(m^\prime_s)$ are the four-momenta (the z-component of spin) of initial proton, $J/\psi$ and final proton, respectively. 
In addition, $u_p$ and $\epsilon^\mu$ denote the spinor and polarization vector of proton and $J/\psi$, respectively. 
Two terms $\mathcal{M}^{\mu \nu}_{\mathbb{P}}$ and $\mathcal{M}^{\mu \nu}_{P_c}$ are the amplitudes for the $\gamma + p \to J/\psi + p$ by Pomeron exchange and the $P_c$ resonances, respectively.

The Pomeron exchange diagram is regarded as background channel. 
The detailed derivation of $\mathcal{M}^{\mu\nu}_\mathbb{P}$ can be found in Ref.~\cite{Wu:2019adv}.
Here we just list them as follows,
    \begin{align*}
    \mathcal{M}^{\mu \nu}_{\mathbb{P}}&= G_{\mathbb{P}}
    \left[ i12\frac{em_{J/\psi}^2}{f_V}\beta_{c}\beta_{u/d}F_{J/\psi}(t)F_1(t)\left(\slashed{q}g^{\mu\nu} - q^\mu \gamma^\nu\right)\right],
        \\
        G_{\mathbb{P}} &=(\frac{W^2}{s_0})^{\alpha_P^\prime t+\alpha_0-1}\exp\{-\frac{i\pi}{2}[\alpha_P^\prime t+\alpha_0-1]\},
        \\
        F_{J/\psi}(t) &= \frac{1}{m_{J/\psi}^2-t}\left(\frac{2\mu_0^2}{2\mu_0^2+m_{J/\psi}^2-t} \right),
        \\
        F_1(t) &= \frac{4m_{N}^2-2.8t}{(4m_{N}^2-t)(1-t/0.71)^2},
    \end{align*}
where $G_{\mathbb{P}}$ is the propagator of Pomeron with $\alpha^\prime_p=1/S_0=0.25$ GeV$^{-2}$ and $\alpha_0=1.25$. 
The $\beta_{u/d/c}$ is the coupling between the Pomeron and quark in the hadron, with $\beta_{u/d}=2.07$ GeV$^{-1}$ and $\beta_{c}=0.84$ GeV$^{-1}$.
Form factor $F_{J/\psi}$ and $F_N$ are for the interaction of Pomeron with $J/\psi$ and $N$ respectively, where $\mu_0=1.1$ GeV and $t=(p-p^\prime)^2$ is in unit of GeV$^2$.

For a signal channel, the amplitude $\mathcal{M}^{\mu \nu}_{P_c}$ is given by the assumption that the spin of $P_c$ is $1/2$ with negative parity. 
\begin{align}
    \mathcal{M}^{\mu\nu}_{P_c} &= \frac{3}{2}g_{1J/\psi}\tilde{\gamma}_\alpha \tilde{g}^{\alpha\mu} \sum\limits_{V=J/\psi,\rho,\omega}F_V(q^2)\frac{ie}{f_V}\frac{-m_V^2\tilde{g}_{1V}}{-m_V^2+i\Gamma_Vm_V}\frac{\slashed{q} + \slashed{p} -m_{P_c}}{W^2-m_{P_c}^2+i\Gamma_{P_c}m_{P_c}}\gamma_\lambda\left(\tilde{g}^{\lambda\nu}-\frac{\tilde{r}^\lambda\tilde{r}^\nu}{\tilde{r}^2}\right),
\end{align}
where $\tilde{g}^{\mu\nu} = g^{\mu\nu} - \frac{(p+q)^\nu (p+q)^\nu}{(p+q)^2}$, $r^\mu =p ^\mu-q^\mu$, $\tilde{\gamma}^\nu = \gamma_{\mu}\tilde{g}^{\mu\nu}$, $\tilde{r}^\nu = r_\mu\tilde{g}^{\mu\nu}$.
$F_V(q^2)$ is the off-shell form factor for intermediate vector as follows,
\begin{align}
    F_V(q^2)=\frac{\Lambda^4_V}{\Lambda^4_V+(q^2-m_V^2)^2},
\end{align}
where cut-off $\Lambda_V$ is undetermined parameter as discussed in Ref.~\cite{Wu:2019adv}.
For simplification, this factor will be deal with a constant number since the main contribution will be around $q^2\sim 0$ GeV$^{2}$ because of the photon propagator.
Here the aim is to find out the kinematic range for largest signal of $P_c$ states, thus, we neglect the interference between the above two mechanisms and the overall constant factors are just taken as 1 for the simplification.
The amplitudes can be calculated by above equations, then the proper set of IKVs is chosen for the best phase space range of the signal of $P_c$ states.
For a three-body final states, there are three sets of IKVs for $e+p\rightarrow e+J/\psi+p$ as shown in Tables.\ref{tab:3a}-\ref{tab:3c}. 
To make the regions minimally overlap between two mechanisms, variables $\ud m^2_{pJ\psi}$,$\ud m^2_{eJ/\psi}$, $\ud\alpha$,$\ud(\cos\beta)$, and $\ud\gamma$ are the most appropriate choices and the z-axis of coordinate frame is along the direction of initial proton.
Typically, the indexes of final states set proton, electron, and $J/\psi$ as particle 1, 2, and 3, respectively, then the physical meaning of Euler angles can be clear and the IKVs set is the same as that in Table.\ref{tab:3a}.
It is worth to mention that $\beta$ can be recognized as the angle between the initial and final protons.
Furthermore, because of the axial symmetry for the scattering process, $\ud\alpha$ can be trivially integrated out and one can get the factor $2\pi$.
On the other hand, $\ud\gamma$ will be integrated through Gaussian quadrature method for the case here because it's hard to be measured directly.
At last, the remaining kinematic variables are $m^2_{pJ/\psi}$, $m^2_{eJ/\psi}$, $\cos\beta$ and the phase space is now three-dimensional.
The distribution of $|\mathcal{M}_{ep\rightarrow eVp}|^2\ud\Phi_{4}/\ud m^2_{pJ/\psi}\ud m^2_{eJ/\psi}\ud \cos\beta$ is straightforwardly computed as a three-dimensional plot with variables $m^2_{pJ/\psi}$, $m^2_{eJ/\psi}$, and $\cos\beta$.

\subsection{Result and discussion}

\begin{figure}[H]
\begin{subfigure}[b]{0.48\textwidth}
{\centering
\includegraphics[width=\textwidth]{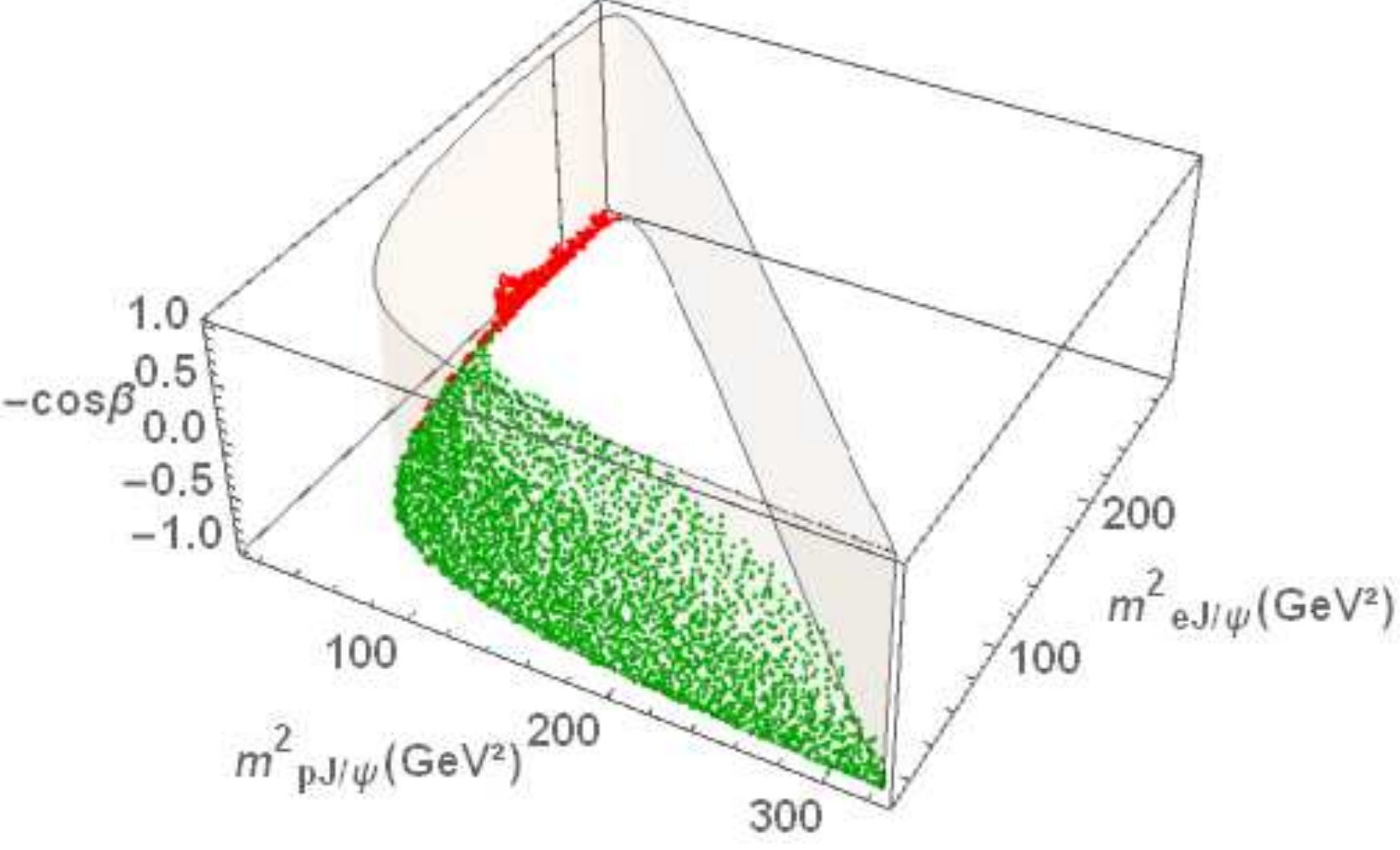}
\caption{}
}
\end{subfigure}
\hfill
\begin{subfigure}[b]{0.48\textwidth}
{\centering
\includegraphics[width=\textwidth]{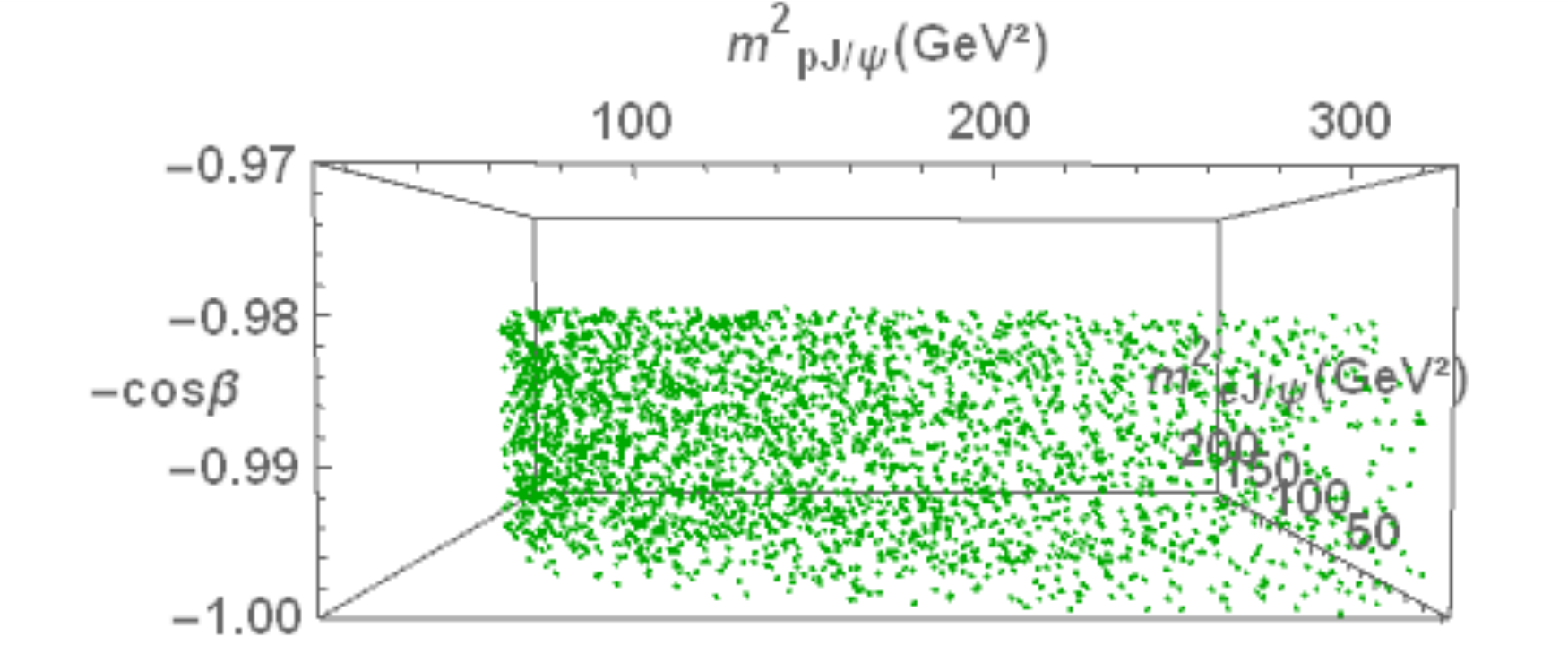}
\caption{}
}
\end{subfigure}
\caption{
Distributions of two channels with the center-of-mass energy being 18 GeV. 
The density of scatter in the neighborhood of a point in the phase space indicates the order of magnitude of the differential cross section at that point.
Regions with no scatter mean that cross section in that region are at least $10^4$ times smaller than the maximum. 
(a) shows the distribution of two channels where red scatter stands for the signal channel while green scatter stands for the background channel. 
(b) shows the distribution of background channel in a detailed scale of $-\cos\beta$}.
\label{fig:res}
\end{figure}

In order to show significant phase space range for two mechanisms, the three-dimensional distribution plots are shown in Fig.\ref{fig:res} with the center-of-mass energy being 18 GeV, which is available for EicC in the future.
Red and green scatters stand for signal channel and background channel, respectively.
The density of scatter in the neighborhood of a point in the phase space indicates the order of magnitude of the differential cross section at that point.
Actually, since there is a huge magnitude difference of the differential cross section for different range, we only draw the main contribution part and leave other blank.   
Obviously, the $P_c$ signals are mainly enriched the edge of plant of two invariant mass variables.
It is easy understood that the signal events will concentrate around the mass of $P_c$ state because of its narrow width.
On the other hand, the background signals concentrate in the range of $\beta \sim 0$ and decrease fast for larger $\beta$.
It results from the exponential term in the $\mathcal{M}_{\mathbb{P}}^{\mu\nu}$.  
Then by this diagram, we find that the best kinematic interval for extracting pentaquark signal for process $e+p\rightarrow e+J/\psi+p$ should require following conditions.
The energy of outgoing electron is larger than 8 GeV, which is calculated from the limitation of invariant mass of $J/\psi p$ system. 
The scattering angle of proton is from 11$\degree$ to 55$\degree$ to avoid the background interference.  
The directions of outgoing proton and outgoing electron are almost anti-parallel because the main contribution are the edge of the Dalitz plot of $m^2_{pJ/\psi}$ and $m^2_{eJ/\psi}$.

\section{A Simple Case for Four-body final states}
\label{sec6}

In the previous three-body case, we find that the resonant peak will be buried in the huge background. 
It is necessary to find a typical kinematic region to figure out the resonance.  
For the four-body case, the possible choices of IKVs are much more than that in the three-body case. 
Thus, it is important to choose a proper set of variables for extracting the information of a typical resonance.

\begin{figure}[htb]
\centering
\includegraphics[scale=0.9]{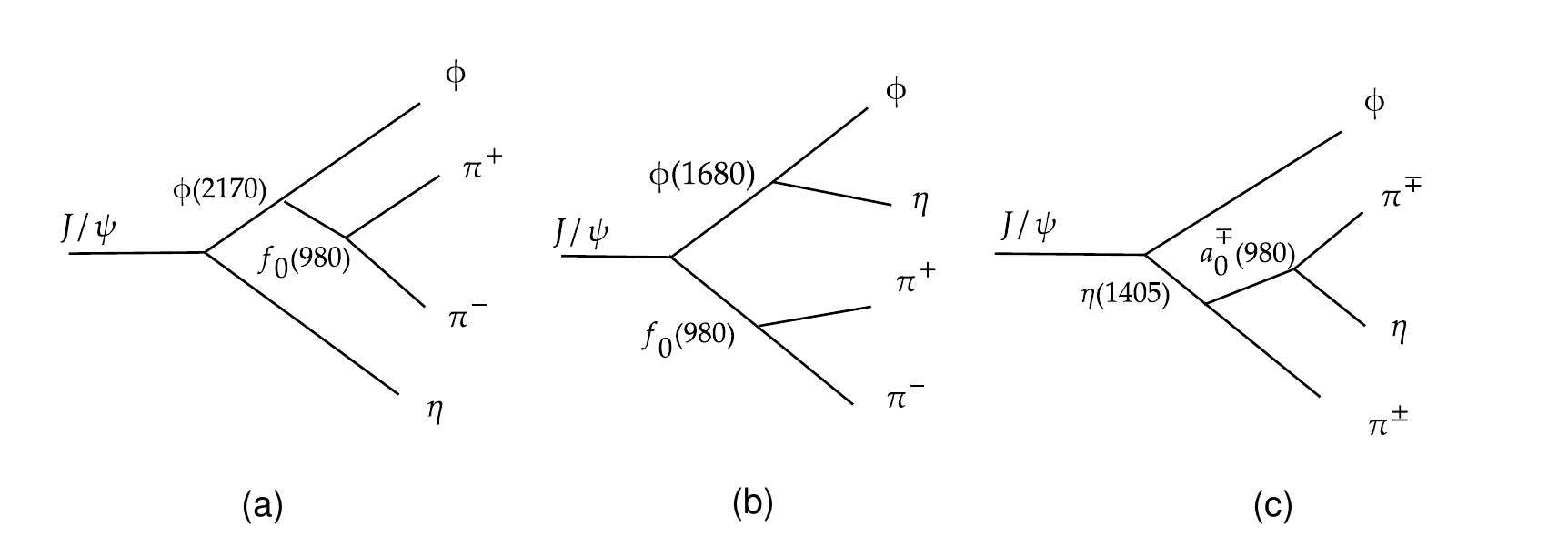}
\caption{Three tree diagrams of reaction $J/\psi \to \eta\pi^+\pi^-\phi $.}
\label{fig:feyJpsi}
\end{figure}

As mentioned before, the invariant mass spectrum is usually useful for extracting the mass and width of the resonance, however, in some cases the resonance peak is not directly seen in the invariant mass spectrum for the reaction with several different mechanisms. 
To illustrate, we provide a toy model for $J/\psi \to \eta\pi^+\pi^-\phi $ reaction here.
Since the aim is just to show the importance of choosing proper kinematic variables, it's simply assumed that the Feynman diagrams for the reaction are just three tree diagrams shown in Fig.~\ref{fig:feyJpsi}. 
For each diagram, there are two resonances. 
To avoid too complicated amplitudes, we just use Breit-Winger propagators of each resonances for a amplitude of each diagram as follows,
\begin{align}
    \mathcal{M} &= \frac{1}{2}A(\phi(2170))A(f_0(980))+ A(\phi(1680))A(f_0(980))+ A(\eta(1405))\left(A(a^+_0(980))+A(a^-_0(980))\right)/2, \label{eq:jpsi}\\
    A(R)&=\frac{1}{p^2_R-m^2_{R}+i\Gamma_R m_R},
\end{align}
where $p_R$, $m_R$, and $\Gamma_R$ are the four-momentum, mass, and width of the resonance $R$.
The mass and width parameters are listed in Table \ref{tab:masswidth}, which are roughly consistent with the values in the RPP.
The additional factor $\frac{1}{2}$ is just used to weaken the contribution of the first diagram.

\begin{table}[htb]
        \centering
\makegapedcells
        \begin{tabular}{ccccccc}
        \hline
		R  &  $\phi(2170)$ & $\phi(1680)$ & $f_0(980)$ & $a^\pm_0(980)$ & $\eta(1405)$ 
            \\
            \hline
 mass   &  $2157$       & $1680$       & $980$      & $980$          & $1405$ \\
 width  &  $100$        & $100$       & $70$       & $70$          & $100$     \\     \hline
\hline
            \end{tabular}
\caption{
The parameters used in Eq.(\ref{eq:jpsi}) with unit MeV.
}
\label{tab:masswidth}
\end{table}

\begin{figure}[htb]
\centering
\includegraphics[scale=0.8]{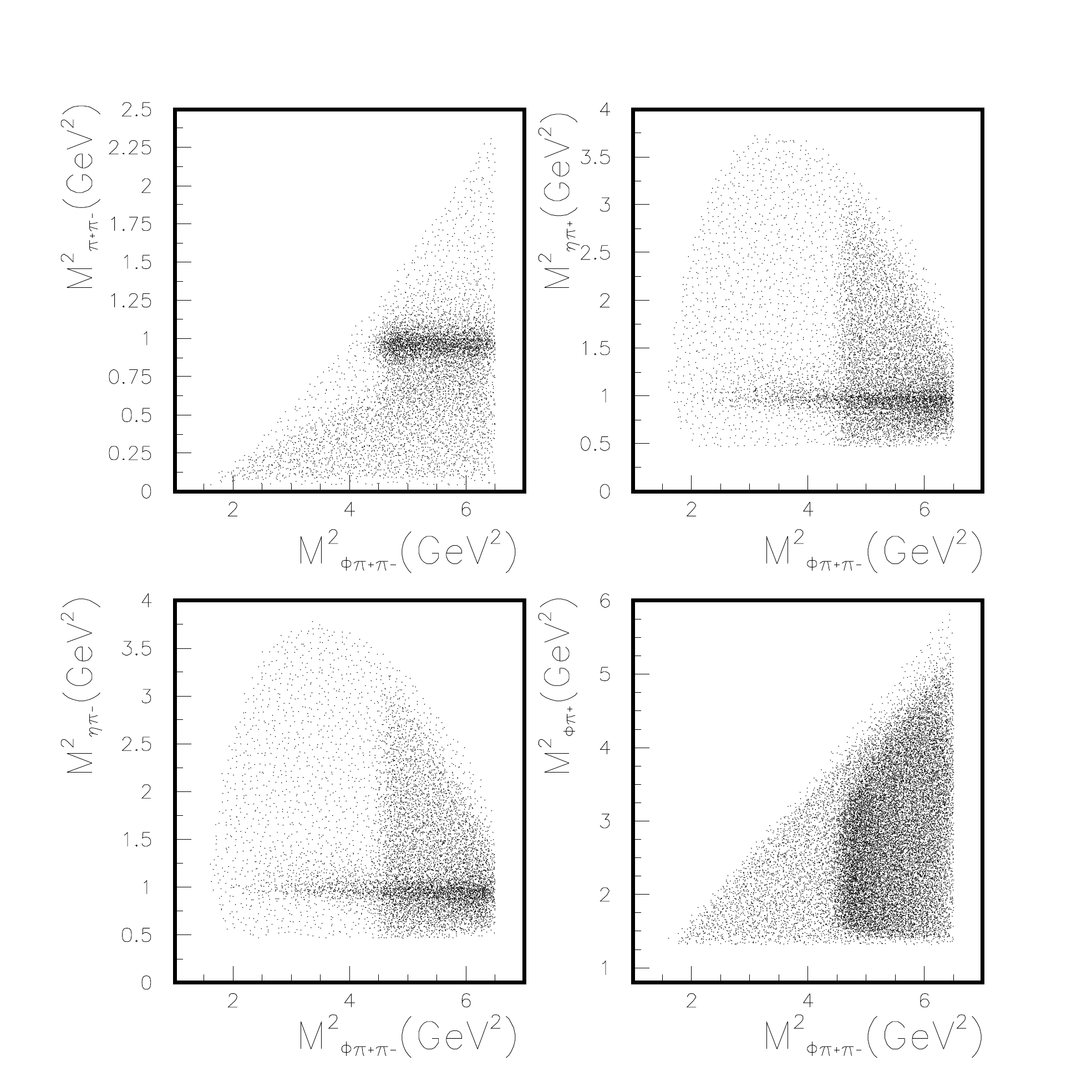}
\caption{The Daliz plots of invariant mass square of $\phi\pi^+\pi^-$ vs various invariant mass squares of $\pi^+\pi^-$, $\eta\pi^+$, $\eta\pi^-$ and $\phi\pi^+$ for reaction $J/\psi \to \eta\pi^+\pi^-\phi $.}
\label{fig:daliz}
\end{figure}

\begin{figure}[htb]
\centering
\includegraphics[scale=0.6]{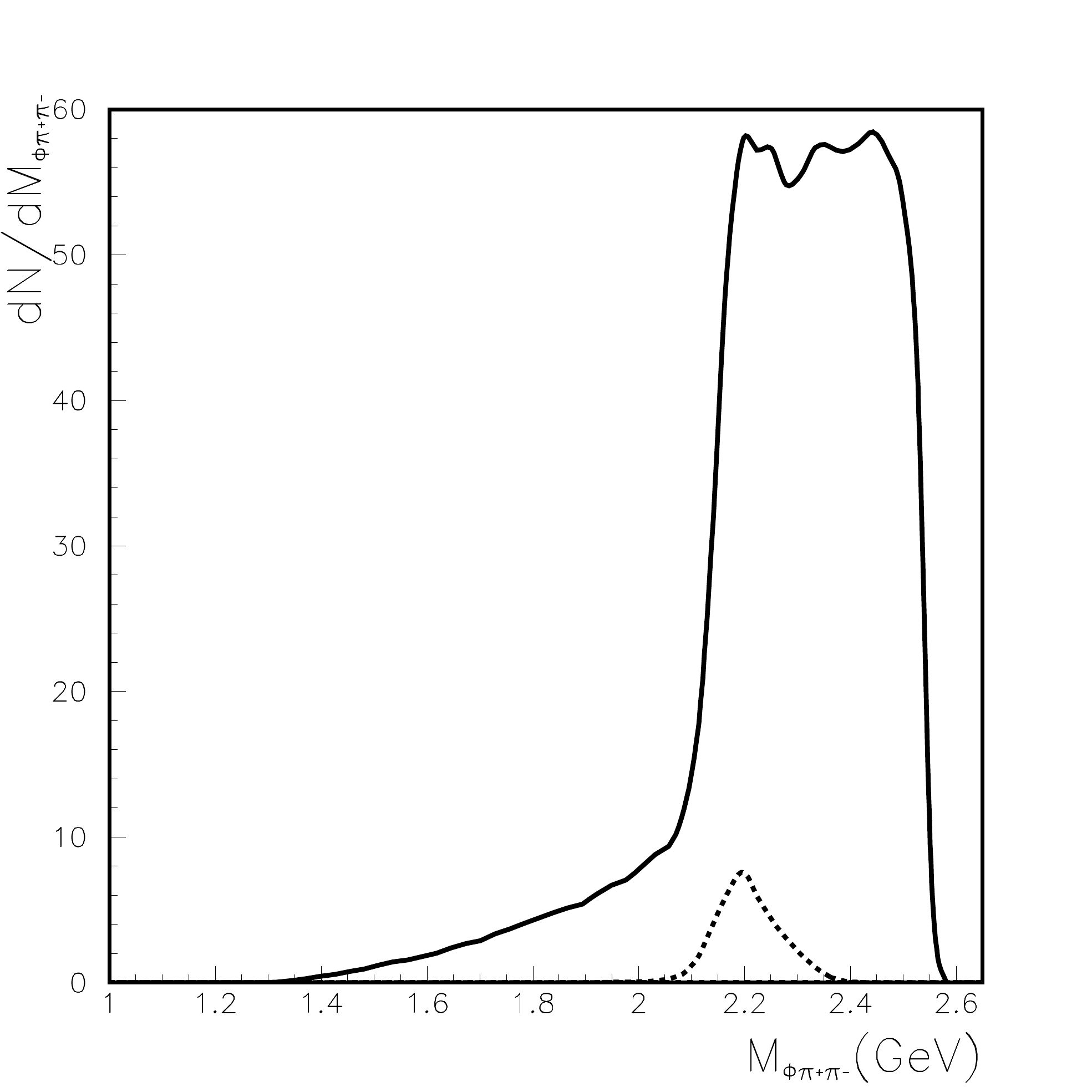}
\caption{The invariant mass spectrum of $\phi\pi^+\pi^-$ for reaction $J/\psi \to \eta\pi^+\pi^-\phi $. $dN/dm_\phi\pi^+\pi^-$ is normalized to the 1000 events of this process. The solid and dashed lines are for the distribution with all contributions and one with a cut $m^2_{\eta\pi^\pm}>1.25$ GeV$^2$, respectively.}
\label{fig:invphipipi}
\end{figure}

Furthermore, we recognize the Fig. \ref{fig:feyJpsi}(a) as a signal channel and the other two mechanisms are both background.
The task is to find a proper way to show the explicit peak of $\phi(2170)$.
Since in this amplitude the variables are all invariant masses, the IKVs shown in Table \ref{tab:4322} are the most convenient choices. Here, $\pi^+$, $\pi^-$, $\eta$, and $\phi$ are recognized as particle $1$, $2$, $3$, and $4$, respectively.
Various Dalitz plots are shown in Fig.\ref{fig:daliz}. 
%
From these Daliz plots, it is clear that the background channels will interfere the peak structure of $\phi(2170)$.
Indeed, from Fig.\ref{fig:invphipipi}, it is found that the typical peak of $\phi(2170)$ resonance is buried in the distribution $\ud N/\ud m^2_{\phi\pi\pi}$ shown in Fig.(\ref{fig:invphipipi}) with black solid line.
On the other hand, from Fig.\ref{fig:daliz}(b)(c) one can easily find that a cut of $m^2_{\eta\pi^\pm}$ is necessary for a clean peak of $\phi(2170)$.
Therefore, with a cut $m^2_{\eta\pi^\pm}>1.25$ GeV$^2$ , a clear peak shown in Fig.(\ref{fig:invphipipi}) with dashed line can be found.
It is a very simple example to show the importance to find a proper kinematic region for extract information of resonance.
In the realistic case, things are much more complicated since the amplitudes can be dependent on various angles by including the high partial wave contributions.
At that time, we believe that this IKVs set including five invariant masses may not a best choice to search the proper kinematic region, other IKVs sets should be useful.

\section{Summary and Prospect}
\label{sec7}
In this paper, all unique sets of kinematic variables containing a certain number of invariant masses are enumerated and classified for three- and four-body final states.
Expressions of phase space factor as well as four-momenta for each case are explicit shown. 
The formulae given in this paper are especially useful for extracting the structure of the resonance.  
As an example of application, we calculate the process $e+p\to e+J/\psi+p$ and find out the region of phase space where the signal and background reach maximum, respectively, which will help experimental physicists to search $P_c$ signal economically and effectively. 
Therefore, the formulae in this paper should be useful for the further researches on three- and four-body final states processes.
Besides, the method provided in this paper, also can be used for any $n$-body final-states processes.
Furthermore, the method developed here can be used for the case that intermediate particle is on shell.
Typically, if it happen in the tree diagram, such intermediate particle can be cut for the final state of former process and initial state for the continue process.
Then whole process can be divided two processes, three- and four-body phase spaces are also applied for it, which is also discussed in Ref.\cite{Heinrich:2008si}.
If it happen in the loop integral, the loop momentum can be recognized as a phase space integration momentum.
So our phase space integration method can also be used in the on-shell loop integration, especially, it may combine with some useful techniques such as sector-decomposition~\cite{Heinrich:2008si}.   
%

\section*{ACKNOWLEDGEMENTS}

The authors would like to thank Hao-Jie Jing, Feng-Kun Guo, Yan-Ping Huang, and Bing-Song Zou
for helpful discussions. 
The work is supported by the Fundamental Research Funds for the Central Universities, 
and by the National Key R$\&$D Program of China under Contract No. 2020YFA0406400,
and by the Key Research Program of the Chinese Academy of Sciences, Grant NO. XDPB15.


	\newpage
	\bibliography{refs}
\end{document}